\documentclass[journal]{IEEEtran}
\usepackage{cite}
\usepackage{tabularx}
\usepackage{graphicx}
\usepackage{subcaption}
\usepackage{mdwmath}
\usepackage{amsmath,amssymb,amsfonts}
\usepackage{mathtools}
\usepackage{epstopdf}
\usepackage{tabularx}
\usepackage{array}
\usepackage{authblk}
\usepackage{cleveref}
\usepackage{soul,color}
\usepackage{booktabs, makecell}

\ifCLASSINFOpdf
\else
\fi

\begin{document}
\title{NAADA: A Noise-Aware Attention Denoising Autoencoder for Dental Panoramic Radiographs}

\author{Khuram~Naveed,~Bruna~Neves~de~Freitas~and~Ruben~Pauwels 
\thanks{K. Naveed, and R. Pauwels are with the Department of Dentistry and Oral Health, Aarhus University, Denmark (knaveed@dent.au.dk, ruben.pauwels@dent.au.dk). B. N. de Freitas is simultaneously affiliated with  Department of Dentistry and Oral Health and Aarhus Institute of Advanced Studies, Aarhus University, Denmark, (brunanf@aias.au.dk).}
\thanks{\textbf{This work has been submitted to the IEEE for possible publication. Copyright may be transferred without notice, after which this version may no longer be accessible.}}
}

\markboth{This article is submitted to the IEEE Transactions on Medical Imaging for possible publication}%
{Shell \MakeLowercase{\textit{et al.}}: Bare Demo of IEEEtran.cls for IEEE Journals}

\maketitle

\begin{abstract}
Convolutional denoising autoencoders (DAEs) are powerful tools for image restoration. However, they inherit a key limitation of convolutional neural networks (CNNs): they tend to recover low-frequency features, such as smooth regions, more effectively than high-frequency details. This leads to the loss of fine details, which is particularly problematic in dental radiographs where preserving subtle anatomical structures is crucial. While self-attention mechanisms can help mitigate this issue by emphasizing important features, conventional attention methods often prioritize features corresponding to cleaner regions and may overlook those obscured by noise. To address this limitation, we propose a noise-aware self-attention method, which allows the model to effectively focus on and recover key features even within noisy regions. Building on this approach, we introduce the noise-aware attention-enhanced denoising autoencoder (NAADA) network for enhancing noisy panoramic dental radiographs. Compared with the recent state of the art (and much heavier) methods like Uformer, MResDNN etc., our method improves the reconstruction of fine details, ensuring better image quality and diagnostic accuracy.
\end{abstract}


\IEEEpeerreviewmaketitle

\section{Introduction}
\label{sec:I}
Dental panoramic radiographs (PR) are x-ray images that capture a broad field of view, including the maxillary and mandibular arches, providing a comprehensive two-dimensional overview of the teeth, jaw, and surrounding structures \cite{farman2007panoramic_chap1}. The ability to visualize these structures in a single image makes PRs an invaluable tool for diagnosing a range of conditions such as impacted teeth, jaw fractures, periodontal diseases, cysts, tumors, bone abnormalities, etc \cite{fuentes2021panoramic}. These properties make it an effective imaging modality for clinical assessment and treatment planning in dentistry. 
However, PRs often suffer from poor image quality due to various sources of noise introduced during the acquisition process \cite{goebel2005noise,arnold1984noise}, which presents a major challenge in its efficacy and clinical utility. Therefore, noise suppression in PRs—both during acquisition and through post-processing—is of pivotal importance.

The prominent granular noise patterns in PRs, known as quantum noise, are primarily caused by low radiation doses, which reduce the signal-to-noise ratio (SNR) and lead to grainy or noisy images.
Apart from that, multiplicative speckle noise, additive white Gaussian noise (AWGN), and impulsive noise (that is, salt and pepper) have been widely observed in the dental radiographs due to various limitations of the acquisition process \cite{abramova2020analysis,goebel2005noise}. In addition, patient movements during exposure and artifacts of dental restorations (such as fillings, crowns, and implants) further degrade the image quality \cite{omar2016quantitative,dhillon2012positioning}. Hence, the resulting mixture of noise often obscures critical diagnostic features such as fine dental structures, carious lesions, periapical pathologies, and bone loss, potentially leading to misinterpretations and suboptimal treatment decisions. Therefore, improving the clarity and diagnostic accuracy of panoramic radiographs is essential to improve clinical outcomes.

In this context, denoising can play a crucial role in improving the quality and diagnostic efficacy of the panoramic radiographs \cite{Abdulbadea2023Enhancing}. To that end, spatial domain filtering, e.g., bilateral filtering \cite{goreke2023novel}, is known to blur fine anatomical details. While transform-domain denoising methods, like wavelet-thresholding \cite{tacs2023application}, can recover finer details, these often lead to artifacts due to the nonlinearity of the operation. To overcome these issues, an advanced image denoising method, named block matching 3D (BM3D) \cite{dabov2007image}, combines collaborative filtering with transform-domain thresholding in a multi-step procedure that is able to reduce noise while preserving fine details. 
However, the BM3D method typically struggles to handle the complex multi-noise scenarios which are often observed in real-world images (e.g., dental radiographs). Despite these shortcomings, BM3D is still regarded a benchmark in most denoising applications \cite{goyal2020bm3d}.

In recent years, deep learning-based methods have emerged as promising solutions for image denoising. For example, in \cite{MLPBM3D2012burger}, a multilayer perceptron (MLP) was shown to outperform the BM3D method for natural image denoising. Consequently, numerous deep learning models $-$ particularly autoencoder (AE), residual network (ResNet), generative adversarial network (GAN) $-$ have been employed for image denoising \cite{DenoisingReview2020tian}. In this regard, a notable invention involves the residual learning approach in \cite{ResLearnDen2017zhang} that trains a network of $D$ instances of a convolutional layer (with batch norm and ReLU operations) to estimate the noise in the input image for its subsequent subtraction. This network, namely deep noise estimation CNN (DnCNN), was recently applied on dental panoramic radiographs for denoising whereby the network was trained and tested for each noise case separately, for example Gaussian or Salt-and-Pepper noise \cite{tacs2023application}. Recently, a modification of DnCNN named MResDNN employs residual convolutional blocks to facilitate the training of a relatively deeper network for enhanced denoising of medical x-ray images \cite{MResDNN2024mittal}. Apart from these, several other variants of residual learning have been developed in different image denoising tasks \cite{hwang2018inception,shi2019hierarchical}.

DAEs have always been a popular architecture for denoising owing to their ability to extract core image features in the latent space that allows them to learn complex representations and reconstruct clean images from noisy inputs \cite{chen2023DAEreview}. However, convolution operations in the conventional DAE architecture limit their local receptive fields. This limitation hinders their ability to capture long-range dependencies and global context, which are essential for accurately reconstructing the variable and diverse anatomical structures found in medical images, such as those in PRs. To address these challenges, attention mechanisms have been integrated into DAEs to expand their receptive fields and focus on key regions of the image \cite{jhamb2018attentive,singh2022attention}. Self-attention (SA) \cite{vaswani2017attention,zhang2019spatialselfattention}, in particular, has shown effectiveness in enhancing feature extraction by allowing the model to consider relationships between distant regions. However, standard self-attention may fail to emphasize important features when they are obscured by high noise levels, as it tends to prioritize regions with clearer correlations to the clean image. Therefore, existing attention DAEs may struggle to suppress noise without losing important details, especially in complex images like dental X-rays.

To overcome this limitation, we propose a noise-aware self-attention (NASA) mechanism that is subsequently utilized at the bottleneck of the proposed DAE which is then trained for enhancing dental PRs. The proposed approach is targeted to improve the model's ability to retain fine anatomical details, improve the quality of the radiograph and produce cleaner, more accurate denoised outputs, particularly in challenging, low-SNR conditions. To that end, our NASA method introduces an additional noise-based attention map that is computed alongside the standard self-attention scores. This ensures that the proposed noise-aware attention DAE (NAADA) network pays appropriate attention to the regions with higher noise levels, enabling the model to capture relevant high-level features even in severely degraded images. This way, the unique noise challenge associated with radiographs, i.e., a combination of various sources of noise, is sought to be addressed to a degree where critical diagnostic features (e.g., finer anatomical structures) are easily discoverable. That ultimately helps clinicians make accurate diagnoses and treatment decisions.

This article is organized as follows: Introduction is presented in Section \ref{sec:I}, followed by the Preliminaries in Section \ref{sec:II} and the Proposed Methodology in Section \ref{sec:III}. That is followed description of results in Section \ref{sec:IV} with a conclusive summary at the end in Section \ref{sec:V}. 
\section{Preliminaries}
\label{sec:II}
\subsection{Self-attention}
\label{sec:IIA}
Attention represents a map of similarity scores between the feature sets of \textit{a query} and \textit{a key}, which is ultimately applied to \textit{the value} or the input data. In self-attention, both the query and key are derived from the same input data. This implies that self-attention calculates the global significance scores for each set of features and applies these scores to the value to compute the attention output \cite{hwang2018inception,shi2019hierarchical}. In principle, the procedure involves computing the correlation of a feature vector with other feature vectors in the matrix, in order to weigh the significance of each feature in the context of the given task. Given that, a query $Q$, key $K$ and value $V$ are derived from the input feature matrix, an attention map is obtained by multiplying the query and the key which is subsequently applied on the value, as follows: 
\begin{equation}\label{Eq01}
    A_t = Softmax \left(\frac{\bar{Q}^T \cdot \bar{K}}{\sqrt{M^{'}}}\right)\bar{V}
\end{equation}
where $M^{'}=\frac{M}{h}$ where $M$ denotes the channel dimensions and $h$ denotes the number of heads in a multi-head attention scenario. Multihead self-attention divides channel dimensions into multiple groups of queries, keys, and values to enhance the model's ability to learn diverse feature relationships and attend to different aspects of the image simultaneously. The softmax operation converts these correlations into a $0-1$ range which serves as a significance score of each feature in the value $V$. This mechanism for emphasizing highly relevant features has been shown to enhance the performance of CNNs across various computer vision tasks \cite{xie2023attentionSegReview,niu2021reviewAttention}.

\subsection{Denoising Autoencoders}
\label{sec:IIB}
A denoising autoencoder (DAE) extends the core functionality of an autoencoder, i.e., to extract key features responsible for representing the key trends in the data, but with the slightly modified goal of learning robust feature representations that are invariant to noise \cite{vincent2008DAEasrobustfeatureextrators,chen2023DAEreview}. That enables reconstruction of a noise-free version of the input data. Hence, DAEs have a dual role, (i) learning meaningful latent representations and (ii) reconstructing the enhanced quality data. This makes them relevant in applications where the quality of input data is compromised, e.g., medical images. 

Typically, DAEs are composed of two main parts: an encoder and a decoder. The encoder is responsible to turn the core image features into a latent representation from input data while the job of the decoder is to reconstruct a clean image based on these features. For this purpose, the mean squared error (MSE) loss function is generally used.

A major limitation of DAEs is their limited receptive field, i.e., DAEs (like many other CNN architectures) are local processing machines which cannot incorporate global dependencies etc. The advantages of incorporating global similarities (among pixels, regions) has already been demonstrated by attention-based transformer architectures \cite{vaswani2017attention}. Hence, use of attention mechanism is often utilized at the bottleneck of the encoder and decoder to address the limitations of the traditional convolutional architectures, i.e., to increase their receptive field etc. This way, relevant features are emphasized by modeling long-range dependencies and incorporating global context etc. Attention-enhanced DAEs provide improved noise suppression and better preservation of the structural details by focusing on the most relevant features of the input data while suppressing irrelevant or noisy regions \cite{jhamb2018attentive,singh2022attention,tihon2021daema}. Despite their efficacy in various other applications, the use of DAEs for enhancing the quality of dental images is scarce in the available literature. To the best of our knowledge, an attention-DAE (ADA) has never been employed to enhance dental radiographs, which are particularly demanding in terms of sharpness preservation.
\section{Methodology}
\label{sec:III}
This section describes our proposed NAADA network whereby a novel noise-aware self-attention (NASA) mechanism is developed for its subsequent use within an autoencoder for denoising panoramic dental radiographs. To that end, we begin by presenting the noise model, followed by an overview of the dental image dataset and its preparation for training and testing, and a description of the proposed approach.

\begin{figure*}
\centering
		\centering
		\includegraphics[width=\textwidth]{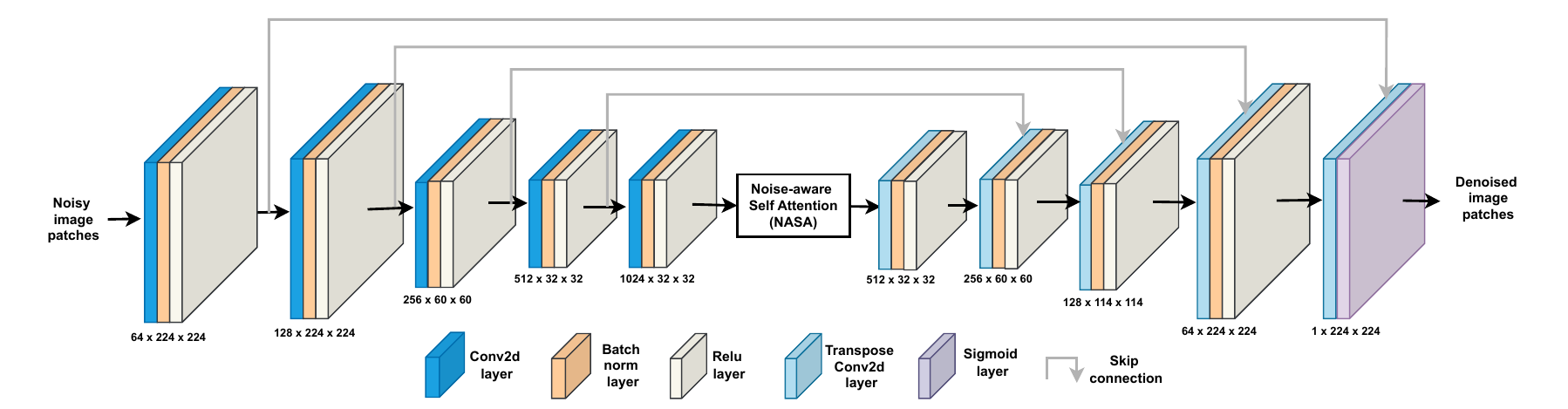}
 \caption{An illustration of the architecture of the proposed NAADA network that employs an encoder to extract core features of the input panoramic image, a novel noise-aware self attention (NASA) mechanism for emphasizing the highly relevant features obscured by noise, and a decoder to get the denoised dental image.}
	\label{fig1}
\end{figure*}
\subsection{Instrumentation and Noise Model}
\label{sec:IIIA}
Dental PRs are subject to diverse sources of noises during acquisition and post-processing. A major source of noise degradation is due to the need to maintain lower radiation levels in order to minimize patient exposure, especially in children and repeat imaging cases \cite{lee2018poisson}. That is because lower radiation doses directly correspond to a reduced signal-to-noise ratio (SNR), resulting in grainy or noisy radiographs. These granular patterns are often described as a combination of different sources of noise, with quantum noise considered the main contributor \cite{kalivas1999modeling}. 
In addition, a multiplicative noise effect, termed speckle, is caused by the interference of the scattered x-rays reaching the detector \cite{abramova2020analysis}.
Moreover, electromagnetic interference, power fluctuations, and transmission inaccuracies can result in dead and saturated pixels, which is termed salt and pepper noise \cite{abramova2020analysis}. 
Hence, for the purpose of training a denoising model on artificially induced noise, we assume a comprehensive, sequential noise model that considers all these scenarios at once in the noisy dental radiographs, as follows
\begin{equation}\label{Eq02}
    Y = \mathcal{S}_n \big ( \mathcal{I}_n \big (\mathcal{G}_n \big ( \mathcal{P}_n \big ( \mathcal{Q}_n \big ( X \big ) \big ) \big ) \big ) \big )
\end{equation}
where $X$ denotes a high-quality dental PR and the symbols $\mathcal{Q}_n (\cdot)$, $\mathcal{P}_n (\cdot)$, $\mathcal{G}_n (\cdot)$, $\mathcal{I}_n (\cdot)$ and $\mathcal{S}_n (\cdot)$ respectively denote the operations introducing Quantum, Poisson, Gaussian, impulsive and speckle noises to yield the noisy image $Y$. The following paragraphs describe each type of noise that was added; for most types, the magnitude of noise was varied within our experiments through consultation with a prosthetic dentist with postgraduate training in radiology, ensuring that the noise patterns were clinically realistic. 

Quantum noise arises from the inherent randomness in the emission and detection of X-ray photons. It is a fundamental component of noise in radiographic imaging due to the detection of fewer photons across the imaging sensors. Hence, in low-dose PRs, the limited photon count at the detector levels leads to more pronounced quantum noise, which manifests as granular, noise-like patterns in the final image.
In this study, we model the quantum noise effect by the Poisson distribution, which accurately represents the probabilistic nature of detected photon counts. We start by simulating the quantum noise, denoted by $\mathcal{Q}_n(\cdot)$ in \eqref{Eq02}, where first step involves the conversion of the gray-scale pixels values in $X$ to the intensity domain, as follows
\begin{equation}\label{Eq03}
    I = 10^{-\frac{X}{c}}
\end{equation}
where $c$ denotes the exposure constant, which was set at $c=50$ \cite{kasap2011amorphous}. Since the intensity values $I$ are directly related to the number of photons $P_{N}$ detected by the imaging sensor, we approximate the photon count by scaling the intensity values by a constant factor, i.e., $P_{N} = \varrho \cdot I$, where $\varrho$ is a scaling constant. To replicate low-exposure imaging conditions, we assume the scaling factor $\varrho$ to be small, i.e., within the range of a few hundreds. Subsequently, we model the resulting photon count as a Poisson process, simulating the stochastic nature of photon arrival, as follows
\begin{equation}\label{Eq04}
    I^{'} = \mathcal{P}_\vartheta  (\varrho \cdot I),
\end{equation}
where $\mathcal{P}_\vartheta(\cdot)$ denotes the Poisson process parameterized by $\vartheta$ that converts the original intensity values $i_n\in I$ to Poisson distributed low exposure intensity values $i_n^{'}\in I^{'}$ where $n = 1, \cdots, N$. 
Subsequently, the resulting intensities are converted back to gray scale image by reversing the aforementioned transformation, as follows:
\begin{equation}\label{Eq05}
    X^{'} = -c \ log_{10}\left({\frac{I^{'}}{\varrho}} \right) 
\end{equation}
where $X^{'}$ denotes gray pixel values for a low-exposure noisy version of $X$ post quantum noise introduction.

We further subject the resulting gray values $X^{'}$ through a Poisson noise process as follows
\begin{equation}\label{Eq06a}
    X^{''} = \mathcal{P}_{\vartheta^{'}}(X^{'}) 
\end{equation}
where $\mathcal{P}_{\vartheta^{'}}(\cdot)$ denotes the Poisson process with parameters $\vartheta^{'}=1$. While Poisson noise added directly to gray values does not represent a distinct physical effect, it was added in this study to represent a combination of exposure reduction, image digitization, and electronic noise.

Next, we model the additive noise component, denoted by $\mathcal{G}_n(\cdot)$ in \eqref{Eq02}, as a zero mean AWGN, as follows
\begin{equation}\label{Eq06b}
    X^{'''} = X^{''} + \mathcal{N}_{\sigma_g}, \ \forall \ n = 1, \cdots, N,
\end{equation}  
where $X^{'''}$ denotes the resulting noisy image and $\mathcal{N}_{\sigma_g}$ denotes the zero mean Gaussian noise with a variance of $\sigma_g^2$ that was selected from a uniform distribution between $0$ and $0.35$.

Subsequently, we account for the interference of the scattered x-rays from the variations in the surfaces of the anatomical structures (like bone and soft tissues) with the primary beam which manifests as a multiplicative speckle noise component, denoted by $\mathcal{S}_n(\cdot)$ in \eqref{Eq02}, as follows
\begin{equation}\label{Eq07a}
    X^{''''} = \mathcal{S}_n(X^{'''}) = X^{'''} +\ X^{'''}\times \mathcal{N}_{\sigma_s}
\end{equation}
where $\mathcal{N}_{\sigma_s}$ denotes zero-mean Gaussian noise of variance $\sigma_s^2$.  
Finally, we introduce the effect of the impulsive salt and pepper noise by randomly turning a fraction of the pixels to maximum (i.e., 255) and minimum (i.e., '0') which we denote as follows
\begin{equation}\label{Eq07b}
    Y = \mathcal{I}_n(X^{''''}).
\end{equation}
In our case we assume that $5\%$ pixels of the input image are affected by salt and pepper noise. 

All together, the different types of noise added provide a challenging task for any denoising model or algorithm, representing all possible sources of image degradation in the radiographic image formation process. 
\subsection{Dataset and Preprocessing}
\label{sec:IIIB}
A common challenge in radiographic image denoising is the lack of high-quality, noise-free dental image data which can serve as reference for the training process. Keeping in mind this consideration, and opting to avoid with purely simulated data, we chose the DENTEX dataset, published in \cite{hamamci2023dentex}, for our work. DENTEX is a publicly released collection of curated PRs aimed to support research in dental image analysis, including tasks such as segmentation and lesion anomaly detection. The images are sourced from routine dental radiographic procedures, offering a range of common dental conditions in high quality, which make it ideal for training denoising algorithms. Specifically, out of a total of $3903$ images in the DENTEX dataset, we selected the $925$ highest-quality images with the help of a clinical expert and subsequently doubled the sample size (i.e., $1850$ images) by making a horizontally mirrored version of each image. The noisy versions are then prepared through the process explained above in Section \ref{sec:IIIB}. The original high-quality image is assumed as a reference (clean or low-noise image) while the noisy version is used as an input for training. 

For training and testing, we divide the dataset in three parts with $70\%$ of the images dedicated to training and $15\%$ each for validation and testing. That means 1296 images are used for training while $277$ are each is used for validation and testing. Since high-resolution images of size $1424\times 2668$ are not efficient for direct training, we divide the selected panoramic radiographs into $91$ patches of size $224\times 224$ each. That makes a training set of $117,936$ image patches and $25,207$ patches for validation. The testing images are patched at the run-time and re-assembled after inference for computing denoising metrics on whole PRs.
\begin{figure*}
\centering
		\centering
		\includegraphics[width=0.8\textwidth]{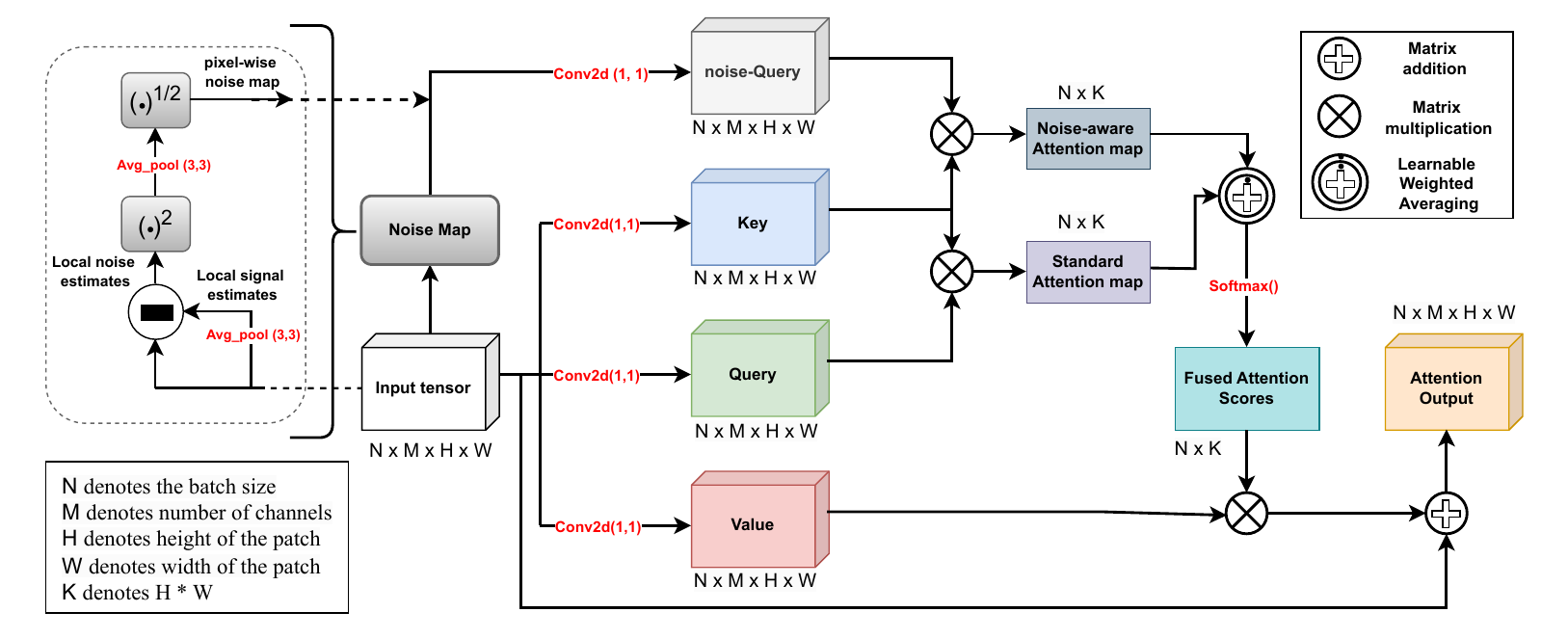}
 \caption{The proposed noise aware self attention (NASA) method that introduces a noise-attention map as a product of the noise-query and the key, apart from the standard self attention map, and subsequently fuses those through weighted averaging to get a noise-aware self attention mechanism.}
	\label{fig2}
\end{figure*}
\subsection{Proposed Noise Aware Attention DAE}
\label{sec:IIIC}
Here, we describe the proposed network architecture designed to denoise panoramic radiographs. Specifically, we develop a novel noise-aware self-attention (NASA) method that is subsequently used within our autoencoder for improved denoising performance. Embedding the noise information in SA makes it more aware of the important features obscured in the noisier regions that allows to extract these features at the decoder during reconstruction. An illustration of the proposed network is shown in Fig. \ref{fig1}; it involves an encoder to extract a feature map from the input image, the proposed noise-aware attention mechanism to emphasize on core image features, and a decoder to reconstruct the denoised image.
The main contributions of this work include: a novel noise-aware self-attention (NASA) method, its use to make a powerful noise-aware attention DAE (NAADA) network that is capable of suppressing noise in dental panoramic radiographs while preserving fine anatomical details. 

In line with the classical denoising autoencoding networks \cite{singh2022attention}, we feed the noisy image patches $\bar{Y}$ to the network while using the corresponding clean patches $\bar{X}$ as labels where the bar $\bar{\cdot}$ over the matrix notation $X$ and $Y$ denotes the multi-dimensionality of the matrices, e.g., a batch of images or an output of a convolution layer etc. 
The encoding block $\bar{F}_{\tilde{\Theta}}=\mathcal{E}_\Theta(\bar{Y},\theta)$ takes a batch of images $\bar{Y}\in [0,1]^{N\times 1\times H\times W}$ as input and returns the feature map $\bar{F}_{\tilde{\Theta}}\in [0,1]^{N\times M \times H^{'}\times W^{'}}$ where $M$ denotes the number of channels and $H^{'}$ and $W^{'}$ denote the updated height and width of the feature map due to repeated down-scaling from the shallow and hidden ($2$D) convolutional layers. In doing so, the encoder network trains its weights $\Theta \in \mathbb{R}$ to extract relevant features required for noise-free reconstruction of the image subject to the appropriate selection of hyper-parameters $\theta$. The role of the noise-assisted attention block $\bar{F}^{'} = \mathcal{A}_{\dot{\Theta}} (F_{\tilde{\Theta}},\theta)$, with learnable weights ${\dot{\Theta}} \in \mathbb{R}$, is to return the enhanced feature map $\bar{F}^{'}\in [0,1]^{N\times M \times H^{'}\times W^{'}}$ such that features relevant to the noise-less image are emphasized. Finally, the decoder block $\hat{\bar{X}}_{\tilde{\Theta}} = \mathcal{D}_{\ddot{\Theta}}(F^{'},\theta)$ reconstructs the batch of images $\hat{\bar{X}}_{\tilde{\Theta}}\in [0,1]^{N\times H\times W}$ such that network weights ${\ddot{\Theta}}\in \mathbb{R}$ are trained to yield an outcome close to the corresponding low-noise batch $\bar{X}_{\tilde{\Theta}}\in [0,1]^{N\times H\times W}$. Here, the symbol ${\tilde{\Theta}}=\{\Theta, \dot{\Theta}, \ddot{\Theta} \}$ denotes the trainable parameters of the whole network. 

The hyperparameters $\theta$ are selected to ensure smooth training of the network weights ${\tilde{\Theta}}$ and minimize the reconstruction error. To that end, we employ the classical mean squared error (MSE) as the batch loss function $L(\bar{X},\hat{\bar{X}}_{\tilde{\Theta},\theta})$, given as follows
\begin{equation}\label{Eq08}
  L(\bar{X},\hat{\bar{X}}_{\tilde{\Theta},\theta})  = \frac{1}{N\times H \times W} \sum_{i=1}^{\tiny{N\times H \times W}} \left(x_i - \hat{x}_i^{\tilde{\Theta},\theta}\right)^2,
\end{equation}
where $x_i\in \bar{X}$ denotes the pixels from the clean image batch (which is used as labels) while $\hat{x}_i^{\tilde{\Theta},\theta} \in \hat{\bar{X}}_{\tilde{\Theta},\theta}$ denotes the pixels of the denoised image batch subject to the trained weights $\tilde{\Theta}$ and hyper parameters $\theta$. In this work, we utilize Adam (Adaptive Moment Estimation) optimizer to minimize the loss \eqref{Eq08} for training our denoising autoencoder. Adam is widely used across different domains of deep learning owing to its ability to use adaptive learning rates for each parameter, while, at the same time, maintaining exponentially decaying averages of the past gradients (first moment estimate) and the square of the gradients (second moment estimate) \cite{kingma2014adam}. To train the proposed network, we use an initial learning rate of $0.01$ and first and second momentum of $0.9$ and $0.99$.

\subsubsection{Encoder}
The purpose of an encoder is to progressively extract higher-level features from within an input gray-scale image using a number of convolutional layer operations in a specific sequence. Keeping in view the complexity of the noise in panoramic radiographs, we choose a deep enough encoder to fully capture the higher-level features of the dental radiograph in order to differentiate between the true image features and the noise. Our \textit{NAADA} encoder contains five convolutional layers (which are followed by a batch normalization and a ReLU operation), as shown on the right side of Fig \ref{fig1}. The encoder begins with a $(3 \times 3)$ convolutional layer to output 64 feature maps while preserving the input resolution. Subsequent layers employ $(4 \times 4)$ convolutions with a stride of 2 to progressively increase the feature depth to $128$, $256$, and $512$ while reducing the resolution to $(114 \times 114)$, $(60 \times 60)$, and $(32 \times 32)$, respectively. A final $(3 \times 3)$ convolutional layer refines the features to produce $1024$ feature maps at a $(32 \times 32)$ resolution. The batch norm layer following each convolution operation is used to stabilize and accelerate training by normalizing the activations, reducing internal covariate shift, and allowing for higher learning rates. 

\subsubsection{Noise-aware self-attention} 
We start by describing the rationale which is followed by the description of the proposed attention method.
\paragraph{Rationale} Despite their powerful denoising framework, DAEs suffer from an inherent shortcoming of CNNs which are known to learn the low-frequency features better than the higher frequencies. This works well in natural image processing in general because most real-world images are composed of smooth regions. On the contrary, noise and artifacts, e.g., grains and patterns etc., are mostly categorized as higher frequencies. Hence, a convolutional DAE often compromises on higher-frequency image details, e.g., edges and corners, and constructs a denoised image by rendering lower-frequency content. However, in the case of dental panoramic radiographs, reconstruction of the higher frequency details (variations in teeth and jaw structures) is critical for accurate diagnosis in clinical application. 
Self-attention provides a powerful tool to address this inherent shortcoming of the convolutional DAEs by (i) increasing the receptive field of the convolution operations and (ii) by emphasizing the important image features. However, existing attention methods, when used in DAEs, may focus on features directly relevant to the clean image while disregarding those trapped in more noisy conditions. 
This important consideration motivates the proposed noise-aware self-attention (NASA) method that seeks to enhance the capabilities of convolutional DAEs when used for denoising of panoramic dental radiographs. That is achieved by assigning higher attention scores to the noisy regions, thus enabling the model to attend to key features trapped in the noisy regions. To that end, we estimate a noise map in the input image patches and construct a noise query to embed the noise-based attention scores in the overall attention matrix, as explained below.
\begin{figure}
     \centering
     \begin{subfigure}[b]{0.24\textwidth}
         \centering
         \includegraphics[width=\textwidth]{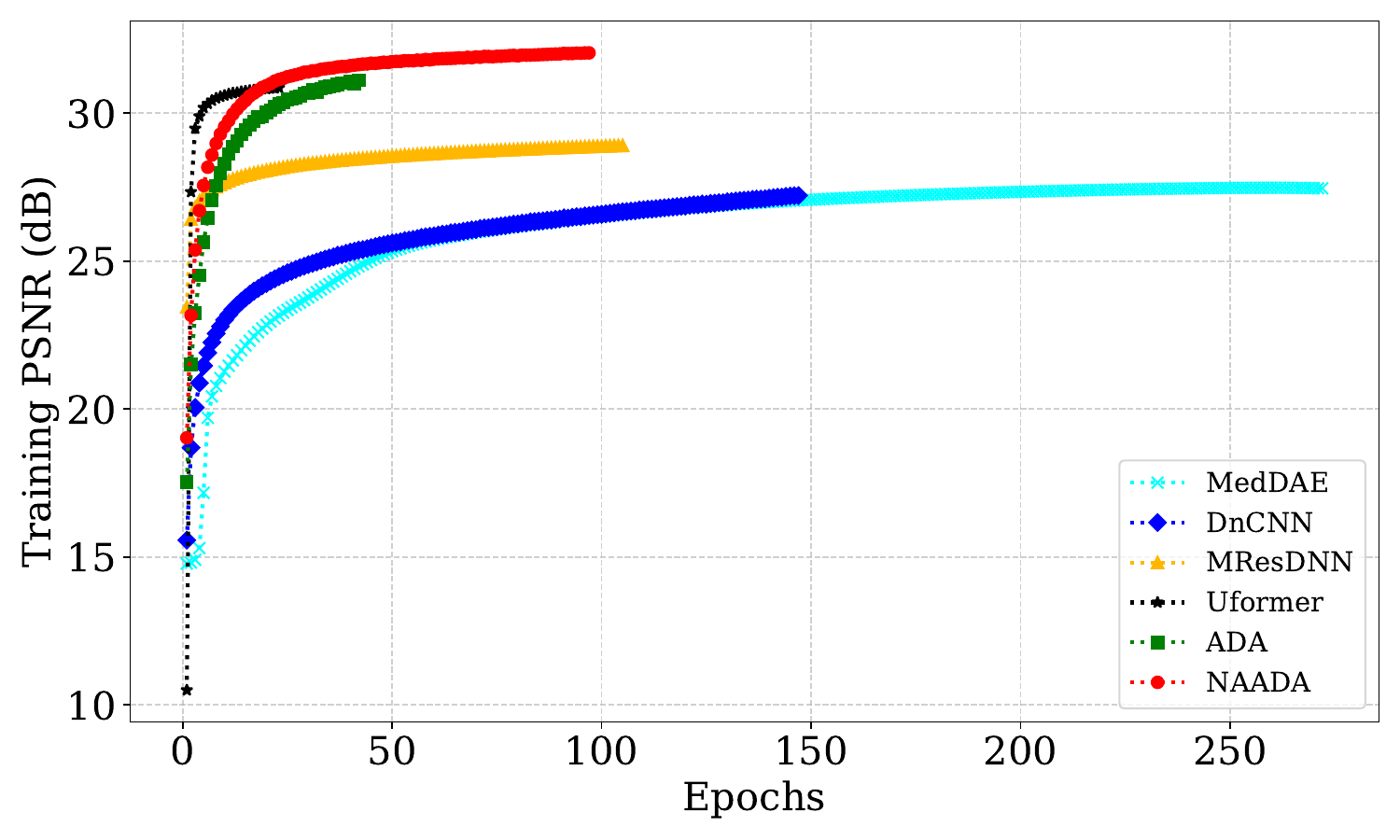}
         \caption{}
         \label{fig3a}
     \end{subfigure}
     \hspace{-2mm}
     \begin{subfigure}[b]{0.24\textwidth}
         \centering
         \includegraphics[width=\textwidth]{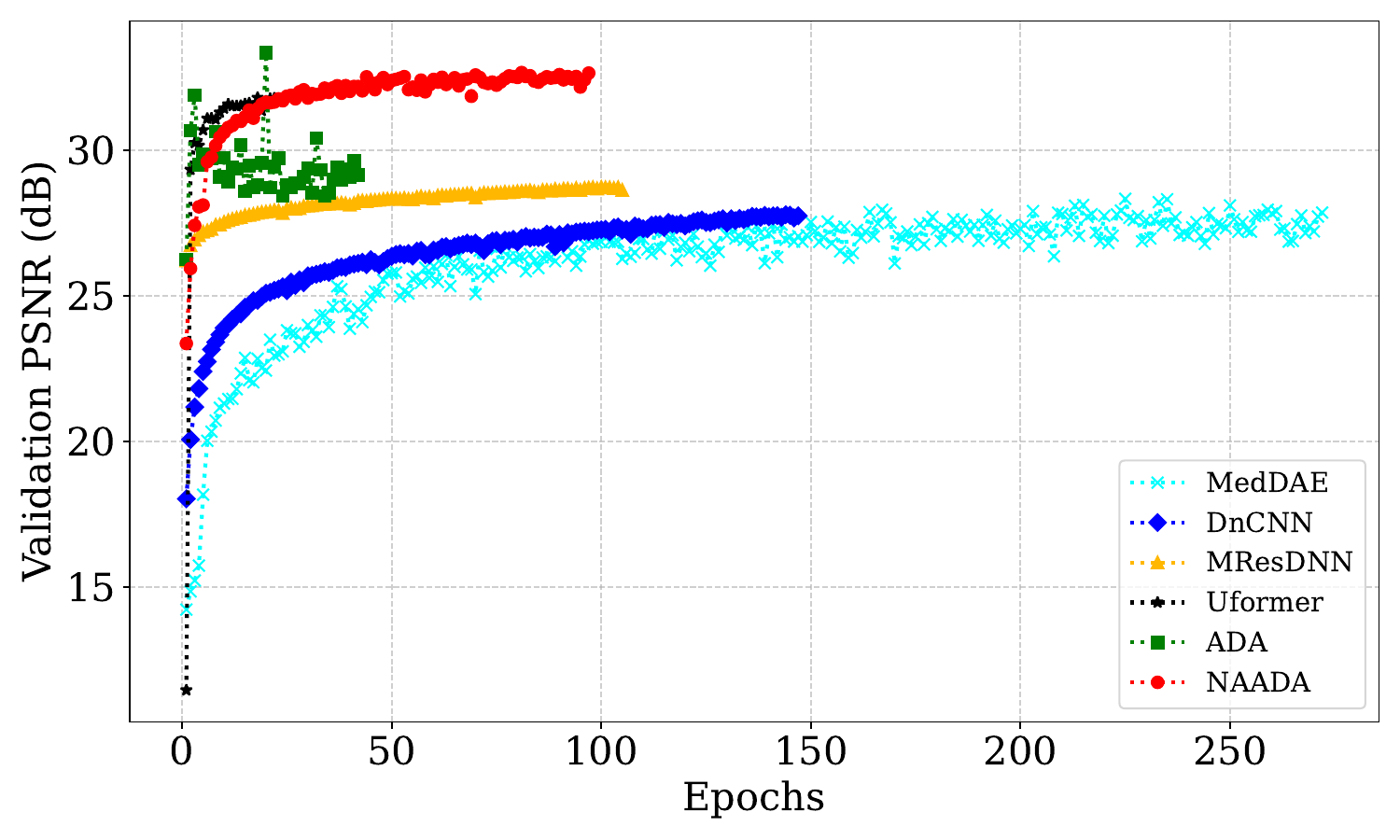}
         \caption{}
         \label{fig3b}
     \end{subfigure}
        \caption{An insight into the training and validation (through the PSNR metrics) of all the deep learning models used in this work including the proposed NAADA and ADA method.}
        \label{fig3}
\end{figure}

\paragraph{Estimating a noise map} Assuming that noise is a phenomenon that introduces artifacts and patterns which present a sharp change from the regular image patterns at the very local level, a local noise estimate $\hat{\psi}_{m_{i,j}}\in \bar{\boldsymbol{\psi}}_{\bar{Z}}$, at location $(i,j)$ and channel $m$ in the noise map $\bar{\boldsymbol{\psi}}_{\bar{Z}}$, can be computed as a root-mean-square (RMS) of the deviation from the local mean, given as follows,
\begin{equation}\label{Eq09a}
    \small \hat{\psi}_{m_{i,j}} =  \sqrt{\frac{1}{k^2}\left(\sum_{i^{'} = i-k^{'}}^{i+}\sum_{j^{'} = j-k^{'}}^{j+k^{'}}\left(z_{m_{i,j}} - \tilde{z}_{m_{i,j}} \right)^2\right)}
\end{equation}
where $z_{m_{i,j}}$ denotes a value of the (feature) in location $(i,j)$ in the $m$\textit{th} channel $Z_m$ of the feature matrix $\bar{Z} \in [0,1]^{N\times M \times H^{'}\times W^{'}}$, that is, $z_{m_{i,j}}\in Z_m\in\bar{Z}$ and the symbol $\small{\tilde{z}_{m_{i,j}}}$ denotes the mean signal value in a small locality, as follows
\begin{equation}\label{Eq09b}
    \small \tilde{z}_{m_{i,j}} =  \frac{1}{k^2}\left(\sum_{i^{'} = i-k^{'}}^{i+k^{'}}\sum_{j^{'} = j-k^{'}}^{j+k^{'}}z_{m_{i^{'},j^{'}}}\right)
\end{equation}
where $k^{'}=\frac{k-1}{2}$ with $k=3$ denotes the kernel or window size for the local estimate of the noise map $\bar{\boldsymbol{\psi}}$. This procedure is depicted in an enclosed box on the right side of Fig. \ref{fig2} where an average pooling layer with a kernel size $k=3$ is used to calculate the local mean. The first average pooling operation $AvgPool_{(3,3)}$ estimates an average signal value $\tilde{z}_{m_{i,j}}$ followed by the computation of the noise estimate while the next average pooling operation is used to estimate the local mean as part of the RMS value computations.
\begin{table}[t!]
    \centering
    \caption{Average PSNR and SSIM results from comparative methods on Test data}
    \label{tab:testing_results}
    \begin{tabular}{l|l|ccc}
        \toprule
         \textbf{Methods} & \textbf{Year}& \textbf{PSNR (dB)} & &\textbf{SSIM} \\
        \hline
        \midrule
         Input &   -& 14.86 && 0.0900 \\
        \hline
        BM3D \cite{dabov2007image} & 2007 & 24.33 ± 0.26 && 0.7191 ± 0.0098 \\
        Med. DAE \cite{gondara2016medical}& 2016 & 27.08 ± 0.23 && 0.6541 ± 0.0082 \\
        DnCNN \cite{ResLearnDen2017zhang} & 2017 & 28.08 ± 0.42 && 0.5750 ± 0.0224 \\
        MResDNN \cite{MResDNN2024mittal} & 2024 & 28.78 ± 0.33 && 0.7084 ± 0.0131 \\
        Uformer \cite{wang2022uformer} & 2021 & 31.35 ± 0.35 && 0.8097 ± 0.0088 \\
        \hline
        Prop. ADA  & 2025& 30.25 ± 0.34  &&  0.7869 ± 0.0096 \\
        Prop. NAADA  &2025&  \textbf{31.86} ± 0.37  &&  \textbf{0.8146} ± 0.0090 \\
        \bottomrule
    \end{tabular}
\end{table}

\paragraph{Noise aware self attention (NASA)} 
Self-attention (SA) is an effective way to help DAEs expand their receptive field and emphasize key features critical to the reconstruction of the denoised image. However, the standard attention mechanism \eqref{Eq01} does not account for the possibility that important features might be obscured by high noise levels. As a result, SA might mistakenly de-emphasize these features as irrelevant, while focusing only on features from less noisy regions due to their clearer correlation with the clean image. 
To address this limitation, we propose a noise-informed self-attention method that computes an additional noise-based attention map alongside the standard self-attention scores. This ensures that the regions with the higher noise levels receive appropriate attention, enabling the model to effectively capture relevant high-level features even in noisy areas. Specifically, we generate sets of queries, keys, and values from the input feature map $\bar{Z}\in \mathbb{R}^{B\times C \times H^{'}\times W^{'}}$ using learnable $1\times 1$ convolutions denoted by $\text{Conv}_{1}(\cdot)$, as follows:
\begin{itemize}
    \item Query: $Q({\bar{Z}})=\text{Conv}_{1\times 1}(\bar{Z})$,
    \item Key: $K{\bar{Z}})=\text{Conv}_{1\times 1}(\bar{Z})$, \item Value: $V({\bar{Z}})=\text{Conv}_{1\times 1}(\bar{Z})$.
\end{itemize}
Additionally, we introduce a noise-informed query derived from the noise map $\bar{\boldsymbol{\psi}}_{\bar{Z}}$, which we call:
\begin{itemize}
    \item Noise-query: $Q(\bar{\boldsymbol{\psi}})=\text{Conv}_{1\times 1}(\bar{\boldsymbol{\psi}}_{\bar{Z}})$.
\end{itemize}
The purpose of the query based on the noise map $\bar{\boldsymbol{\psi}}_{\bar{Z}}$ \eqref{Eq09a} is to inform the attention mechanism about noise levels, i.e., giving sufficient weight to the features trapped in the noisy regions. The proposed noise-informed self-attention is defined as follows
\begin{equation}\label{Eq10}
   \small{A_t = Softmax \left(\frac{{Q({\bar{Z}})}^T \cdot K({\bar{Z}}) + \gamma^{'} {Q({\bar{\boldsymbol{\psi}}})}^{T}\cdot K({\bar{Z}})}{\sqrt{M^{'}}}\right)^T\cdot V({\bar{Z}})}
\end{equation}
where $\cdot^T$ denotes the transpose or reshape operator and $\gamma^{'}$ denotes a learnable weight that decides what ratio of noise attention scores needs to be added to the standard attention scores for a desirable denoising performance.

Since Eq. \eqref{Eq09a} defines noise as sharp granular patterns superimposed over relatively smoother image details at the local level, the noise map consists of both regions with higher granular noise patterns and high-frequency image features, such as edges and corners of anatomical structures (e.g., the jaws and teeth). This implies that the term ${Q({\bar{\boldsymbol{\psi}}})}^T \cdot K({\bar{Z})}$ computes an attention map that highlights important image features hidden within noise-corrupted regions in $\bar{Z}$.
Thus, the proposed noise-attention map reinforces the emphasis on essential anatomical structures in noisy regions, complementing the standard self-attention map 
${Q({\bar{Z}})}^T \cdot K({\bar{Z}})$, that primarily highlights features based on their global relevance. That means, important details obscured within the higher noise regions get emphasized along with the standard attention map of prominent features, resulting in enhanced denoising.
\begin{figure}
     \centering
     \begin{subfigure}[b]{0.23\textwidth}
         \centering
         \includegraphics[width=\textwidth]{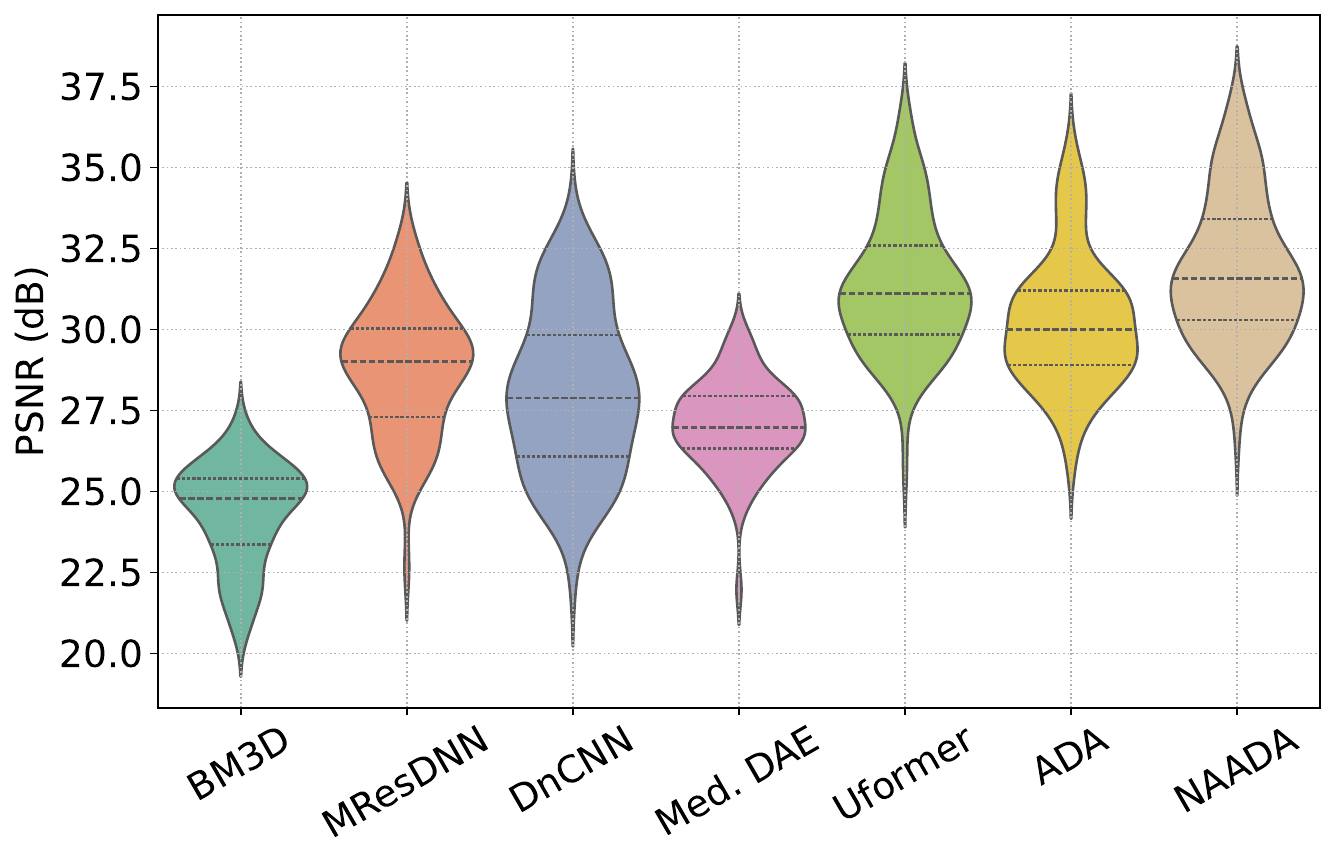}
         \caption{}
         \label{fig4a}
     \end{subfigure}
     \hspace{-1mm}
     \begin{subfigure}[b]{0.23\textwidth}
         \centering
         \includegraphics[width=\textwidth]{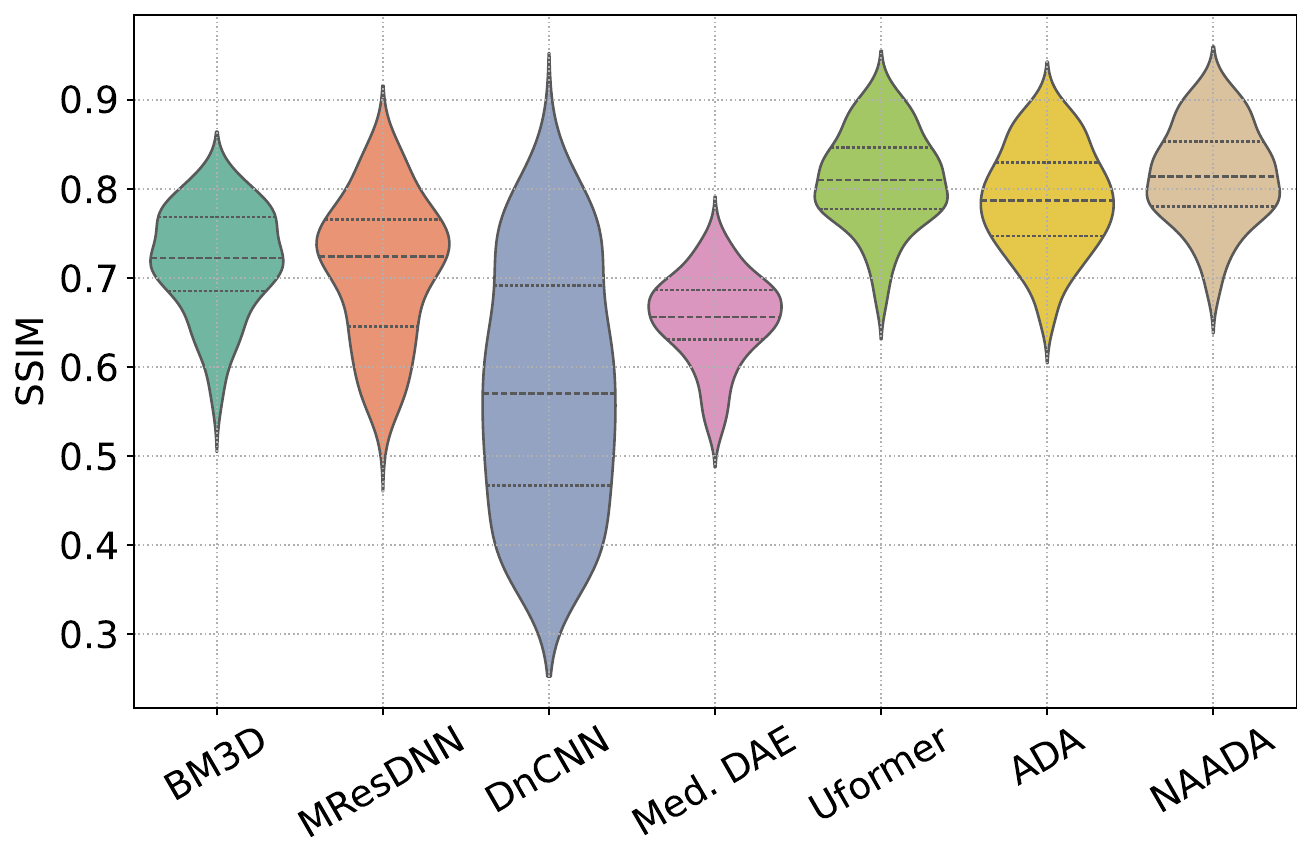}
         \caption{}
         \label{fig4b}
     \end{subfigure}
        \caption{Performance range of comparative methods for the testing data}
        \label{fig4}
\end{figure}

To enhance the model’s ability to capture different types of dependencies, we employ a multi-head attention setup whereby feature dimensions are grouped into multiple heads that allows each head to learn distinct attention patterns. That essentially means that different heads focus on different spatial patterns and noise characteristics. For example, some heads may prioritize the locally relevant high-frequency patterns obscured by noise, while others emphasize anatomical structures or broader contextual relationships within the image. After computing attention in each head, the outputs are concatenated and passed through a Conv2D projection layer to aggregate the information from different heads into a unified feature representation, ensuring that diverse attention patterns contribute to the final feature map. The proposed multi-head noise-aware attention approach strengthens the ability of the DAE to learn diverse and more meaningful structures in the presence of noise, leading to effective denoising and detail preservation.
\subsubsection{Decoder}
The decoder in the proposed NAADA network aims to reconstruct the denoised image from the high-level feature representations extracted by the encoder and refined by our NASA mechanism. To that end, a series of transposed convolutional layers are employed to progressively up-sample the feature maps such that the spatial resolution is restored and fine structural details are accounted for. An illustration of the proposed decoder architecture is presented on the left side of Fig. \ref{fig1} where five transposed convolutional layers are used to mirror the architecture of the proposed encoder, each followed by batch normalization and a ReLU activation function, except for the final layer where a sigmoid activation is employed. Additionally, we employ \textit{skip connections} between the corresponding encoder and decoder layers for ensuring direct transfer of low-level spatial features to preserve fine anatomical structures and mitigate information loss.
\begin{figure}
     \centering
     \begin{subfigure}[b]{0.16\textwidth}
         \centering
         \includegraphics[width=\textwidth]{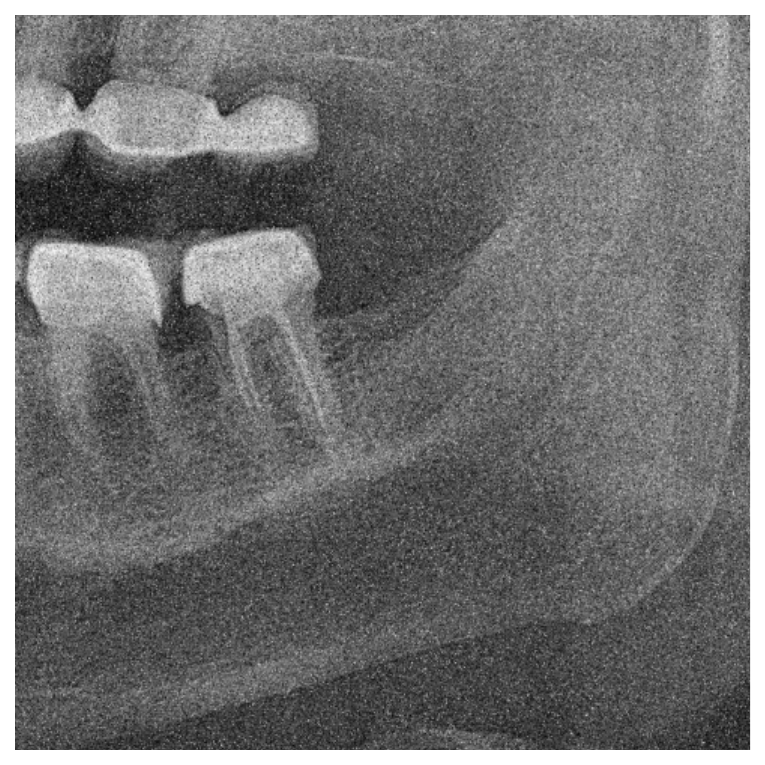}
         \caption{Noisy image}
         \label{fig5a}
     \end{subfigure}
     \hspace{-3mm}
     \begin{subfigure}[b]{0.16\textwidth}
         \centering
         \includegraphics[width=\textwidth]{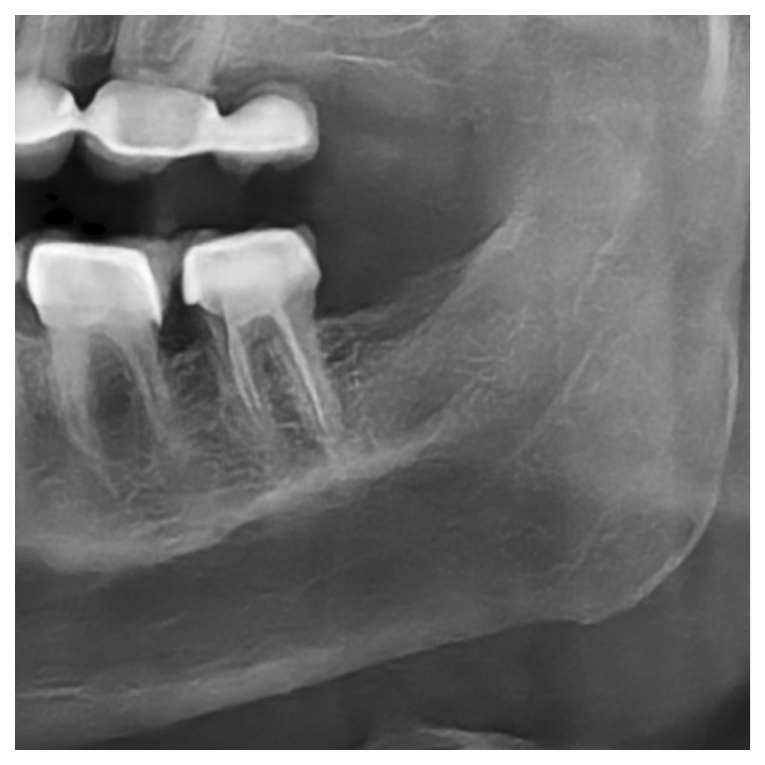}
         \caption{ADA method}
         \label{fig5b}
     \end{subfigure}
     \hspace{-3mm}
     \begin{subfigure}[b]{0.16\textwidth}
         \centering
         \includegraphics[width=\textwidth]{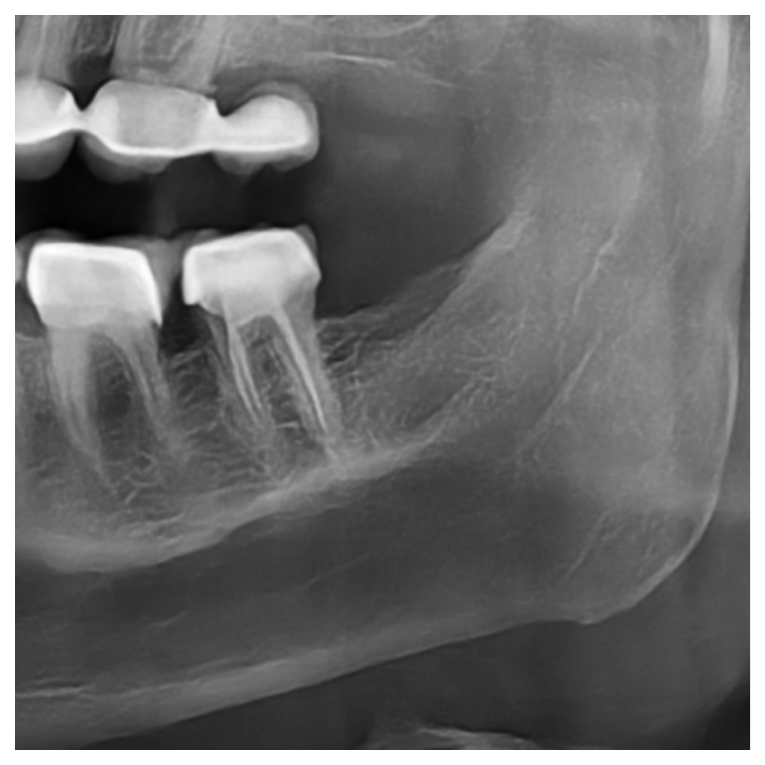}
         \caption{NAADA method}
         \label{fig5c}
     \end{subfigure}
        \caption{Example of the impact of the proposed noise-aware attention method on the denoising performance of the  results obtained through proposed ADA and NAADA methods.}
        \label{fig5}
\end{figure}

The decoder begins with a $(3 \times 3)$ transposed convolutional layer that maintains a spatial resolution of $(32 \times 32)$ while reducing the feature depth from 1024 to 512. Next, a $(4 \times 4)$ transposed convolutional layer with a stride of 2 and padding of 3 increases the resolution to $(60 \times 60)$ and reduces the feature depth to 256. This upsampling process continues with another $(4 \times 4)$ transposed convolutional layer, which increases the resolution to $(114 \times 114)$ and further reduces the feature depth to 128. Following this, a $(4 \times 4)$ transposed convolutional layer upsamples the feature maps to $(224 \times 224)$ with a feature depth of 64. Finally, a $(3 \times 3)$ transposed convolutional layer outputs a single-channel grayscale image of size $(224 \times 224)$ where a sigmoid activation function ensures the pixel values are constrained within the range $[0,1]$.
\begin{figure*}
     \centering
  
     \begin{subfigure}[b]{0.32\textwidth}
         \centering
         \includegraphics[width=\textwidth]{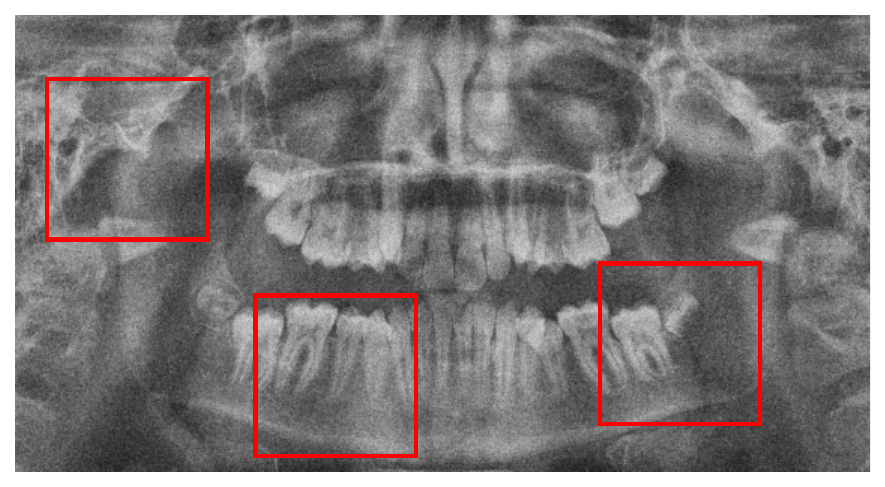}
         % \caption{$y=3\sin x$}
         % \label{fig:three sin x}
     \end{subfigure}
     \begin{subfigure}[b]{0.32\textwidth}
         \centering
         \includegraphics[width=\textwidth]{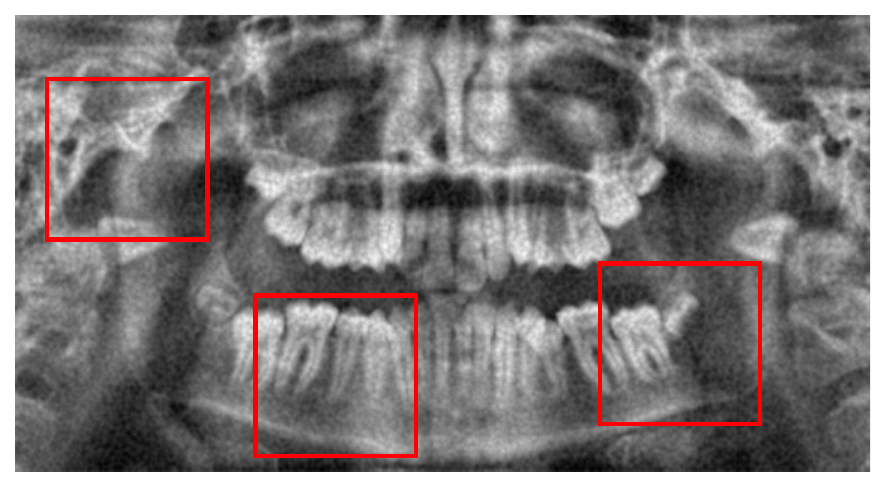}
         % \caption{$y=3\sin x$}
         % \label{fig:three sin x}
     \end{subfigure}
     \begin{subfigure}[b]{0.32\textwidth}
         \centering
         \includegraphics[width=\textwidth]{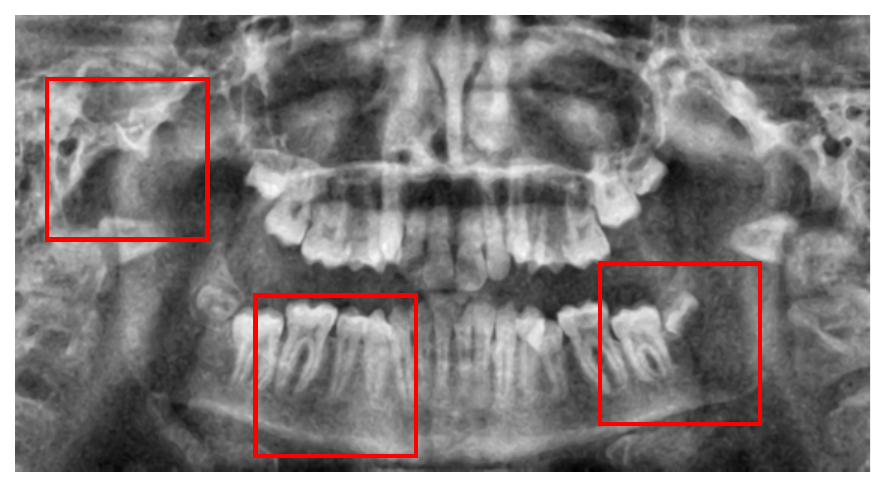}
         % \caption{$y=3\sin x$}
         % \label{fig:three sin x}
     \end{subfigure}
     %\hspace{-6mm}
     
     \begin{subfigure}[b]{0.32\textwidth}
         \centering
         \includegraphics[width=\textwidth]{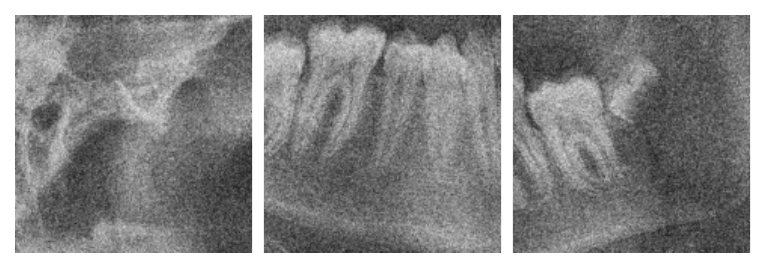}
         \caption{Noisy}
         \label{fig6a}
     \end{subfigure}
     \begin{subfigure}[b]{0.32\textwidth}
         \centering
         \includegraphics[width=\textwidth]{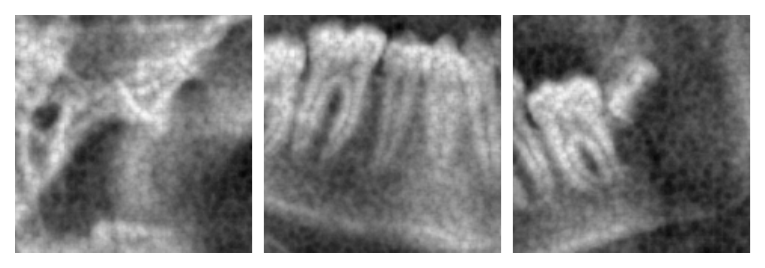}
         \caption{MedDAE}
         \label{fig6b}
     \end{subfigure}
     \begin{subfigure}[b]{0.32\textwidth}
         \centering
         \includegraphics[width=\textwidth]{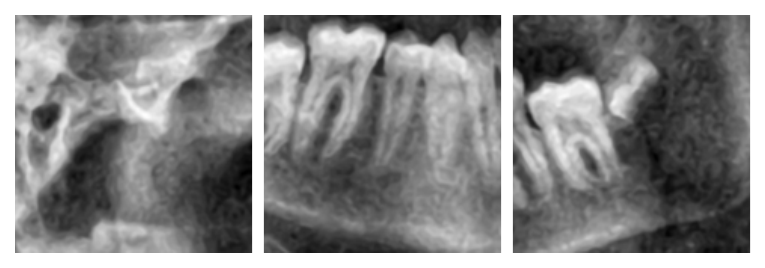}
         \caption{BM3D}
         \label{fig6c}
     \end{subfigure}    
    
     \begin{subfigure}[b]{0.32\textwidth}
         \centering
         \includegraphics[width=\textwidth]{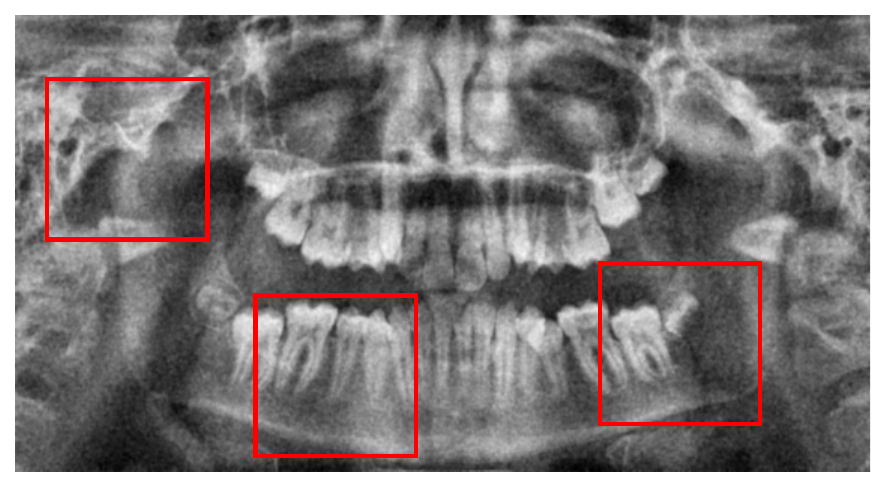}
         % \caption{$y=3\sin x$}
         % \label{fig:three sin x}
     \end{subfigure}
     \begin{subfigure}[b]{0.32\textwidth}
         \centering
         \includegraphics[width=\textwidth]{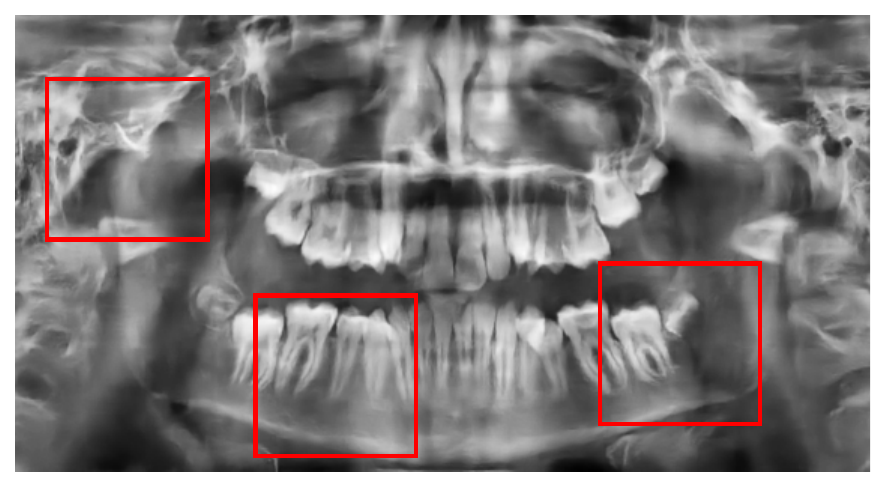}
         % \caption{$y=3\sin x$}
         % \label{fig:three sin x}
     \end{subfigure}
     \begin{subfigure}[b]{0.32\textwidth}
         \centering
         \includegraphics[width=\textwidth]{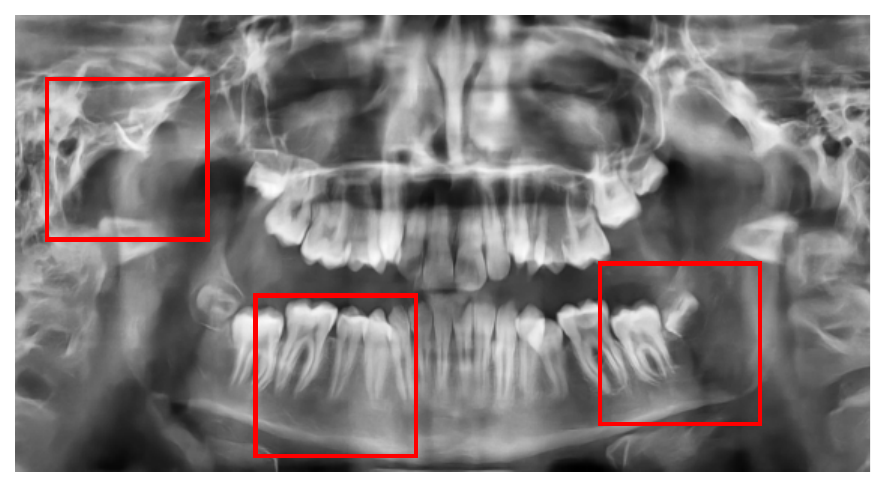}
         % \caption{$y=3\sin x$}
         % \label{fig:three sin x}
     \end{subfigure}

    \begin{subfigure}[b]{0.32\textwidth}
         \centering
         \includegraphics[width=\textwidth]{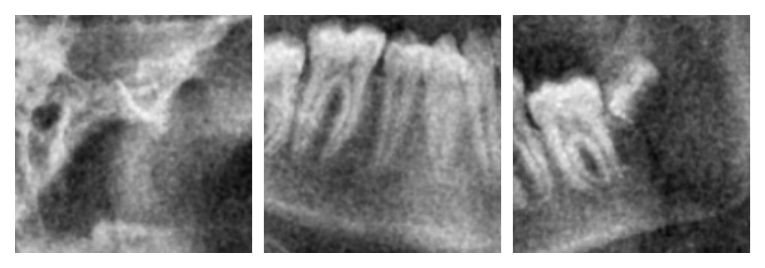}
         \caption{MResDNN}
         \label{fig6d}
     \end{subfigure}
     \begin{subfigure}[b]{0.32\textwidth}
         \centering
         \includegraphics[width=\textwidth]{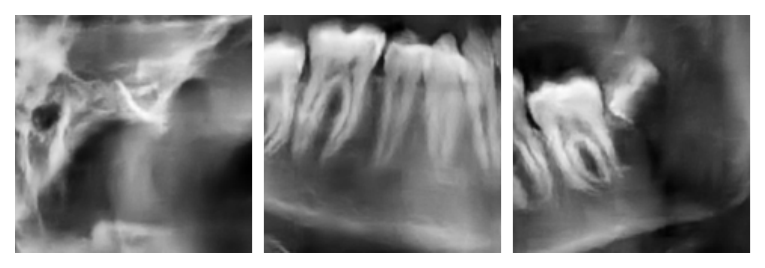}
         \caption{Uformer}
         \label{fig6e}
     \end{subfigure}
     \begin{subfigure}[b]{0.32\textwidth}
         \centering
         \includegraphics[width=\textwidth]{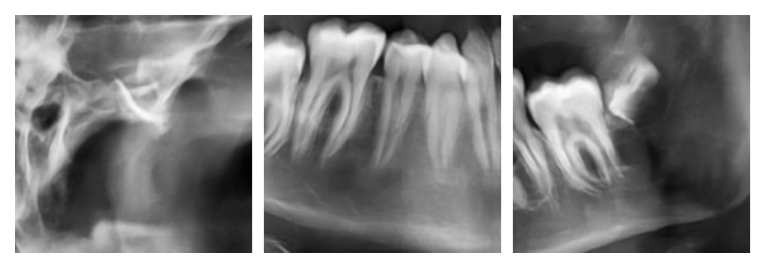}
         \caption{Prop. NAADA}
         \label{fig6f}
     \end{subfigure}
        \caption{Denoising results of the best comparative methods on a test image at input PSNR of $11.74$ dB}
        \label{fig6}
\end{figure*}

\section{Description of Results}
\label{sec:IV}
This section presents the evaluation of the denoising performance of the proposed NAADA method when compared to the state-of-the-art (SOTA) methods which include BM3D, Med. DAE, DnCNN, MResDNN and Uformer methods. Apart from the proposed NAADA and ADA methods, we have implemented and trained Med. DAE, DnCNN, MResDNN networks ourselves based on the description of the models and hyperparameters in their respective articles \cite{dabov2007image, ResLearnDen2017zhang, MResDNN2024mittal, wang2022uformer}. All these networks are trained on a part of the DENTEX panoramic dataset described in Section \ref{sec:IIIB}. To prevent overfitting, an early stopping criterion was implemented by monitoring the validation loss at each epoch. Training was halted when no further improvement in validation loss was observed. We use peak-signal-to-noise-ratio (PSNR) and structural similarity index measure (SSIM) for evaluation of the denoising performance. PSNR measures the quality of the reconstructed image as a ratio between the maximum power of an image and the power of the noise affecting the image. 
SSIM evaluates the structural similarity between the clean and the denoised image, with values closer to 1 indicating better preservation of structural details of the input image. Presented results include plots of training and validation performance metrics, average PSNR and SSIM values of the each method on the testing dataset, violin plots to demonstrate the distribution of the testing performance, and visual results to demonstrate the effectiveness of the proposed method.
\subsection{Quantitative evaluation}
The training and validation PSNR values of the denoised image batches were monitored at each epoch of the training process for all of the deep learning based methods including the proposed NAADA and ADA methods, as shown in Fig. \ref{fig3}.  
\begin{figure*}
     \centering
  
     \begin{subfigure}[b]{0.32\textwidth}
         \centering
         \includegraphics[width=\textwidth]{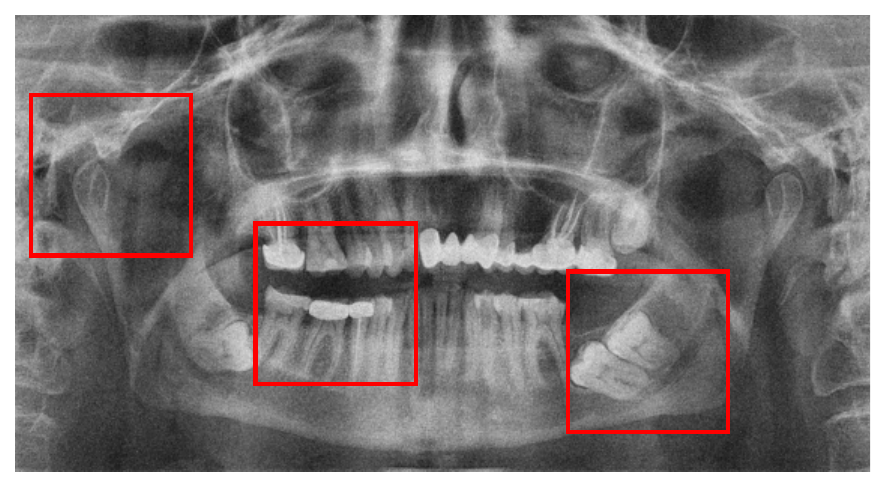}
         % \caption{$y=3\sin x$}
         % \label{fig:three sin x}
     \end{subfigure}
     \begin{subfigure}[b]{0.32\textwidth}
         \centering
         \includegraphics[width=\textwidth]{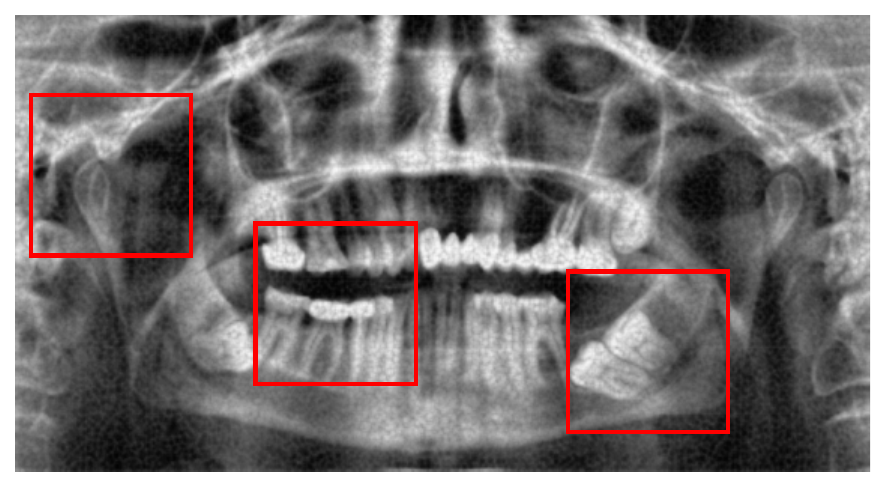}
         % \caption{$y=3\sin x$}
         % \label{fig:three sin x}
     \end{subfigure}
     \begin{subfigure}[b]{0.32\textwidth}
         \centering
         \includegraphics[width=\textwidth]{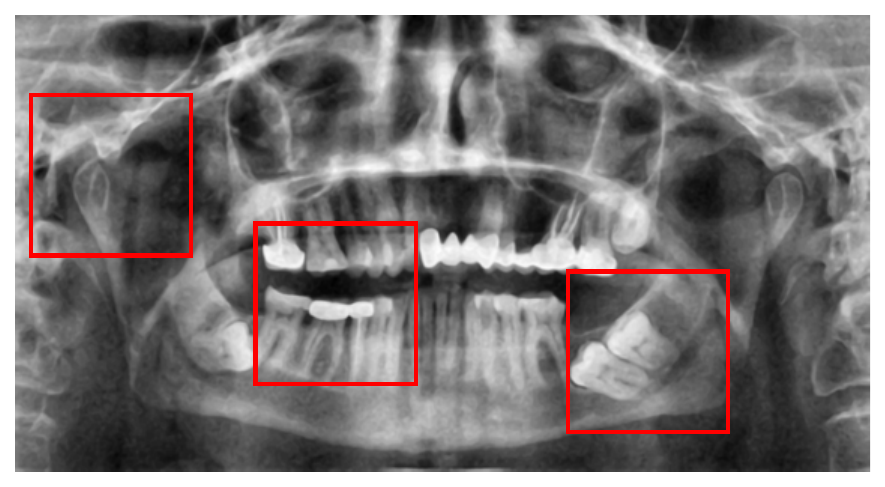}
         % \caption{$y=3\sin x$}
         % \label{fig:three sin x}
     \end{subfigure}
     %\hspace{-6mm}
     
     \begin{subfigure}[b]{0.32\textwidth}
         \centering
         \includegraphics[width=\textwidth]{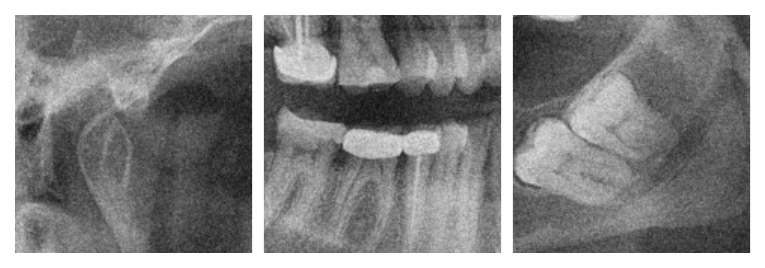}
         \caption{Noisy image}
         \label{fig7a}
     \end{subfigure}
     \begin{subfigure}[b]{0.32\textwidth}
         \centering
         \includegraphics[width=\textwidth]{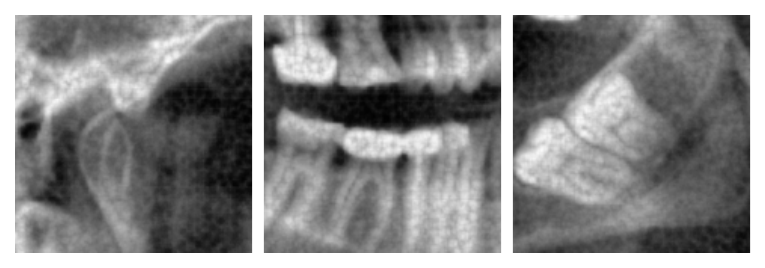}
         \caption{MedDAE method}
         \label{fig7b}
     \end{subfigure}
     \begin{subfigure}[b]{0.32\textwidth}
         \centering
         \includegraphics[width=\textwidth]{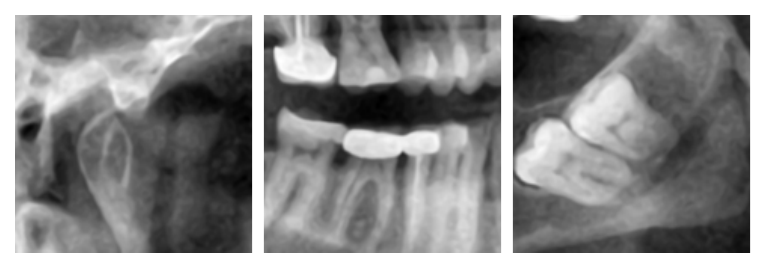}
         \caption{BM3D method}
         \label{fig7c}
     \end{subfigure}
     
    \begin{subfigure}[b]{0.32\textwidth}
         \centering
         \includegraphics[width=\textwidth]{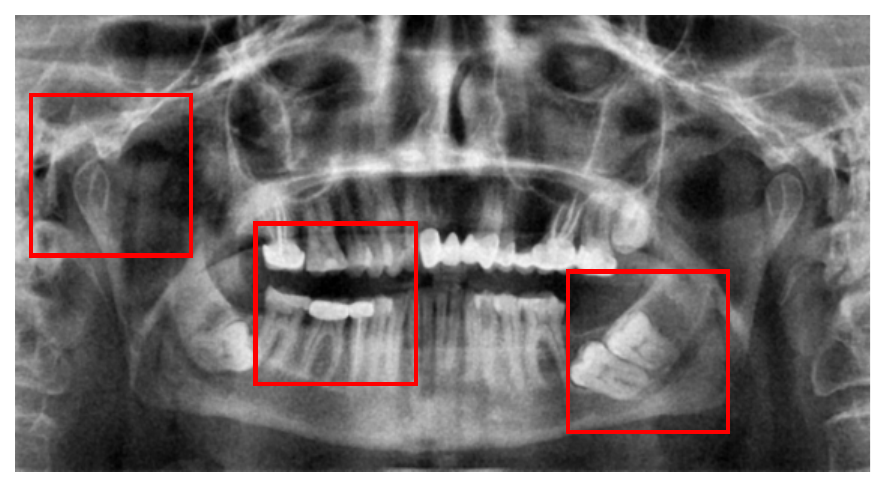}
         % \caption{$y=3\sin x$}
         % \label{fig:three sin x}
     \end{subfigure}
     \begin{subfigure}[b]{0.32\textwidth}
         \centering
         \includegraphics[width=\textwidth]{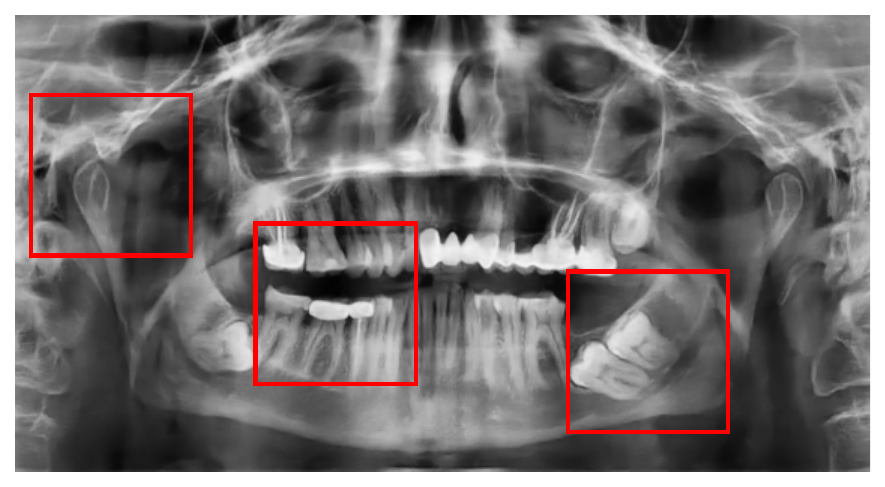}
         % \caption{$y=3\sin x$}
         % \label{fig:three sin x}
     \end{subfigure}
     \begin{subfigure}[b]{0.32\textwidth}
         \centering
         \includegraphics[width=\textwidth]{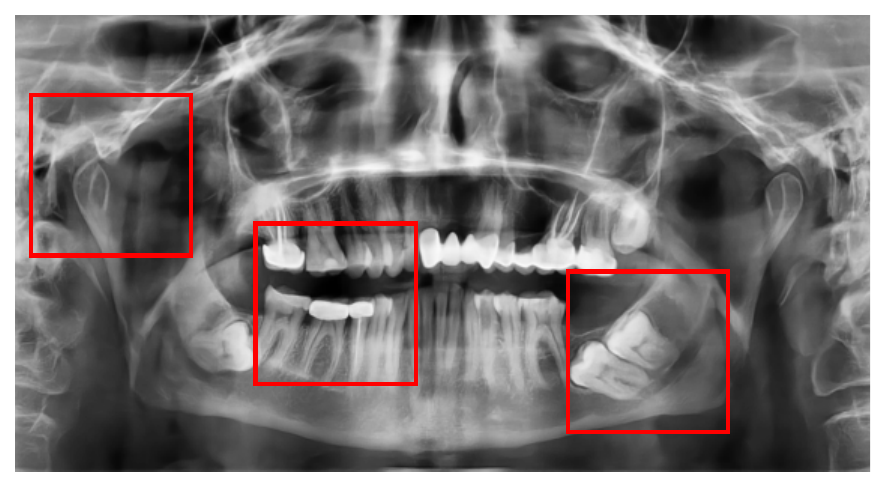}
         % \caption{$y=3\sin x$}
         % \label{fig:three sin x}
     \end{subfigure}

     \begin{subfigure}[b]{0.32\textwidth}
         \centering
         \includegraphics[width=\textwidth]{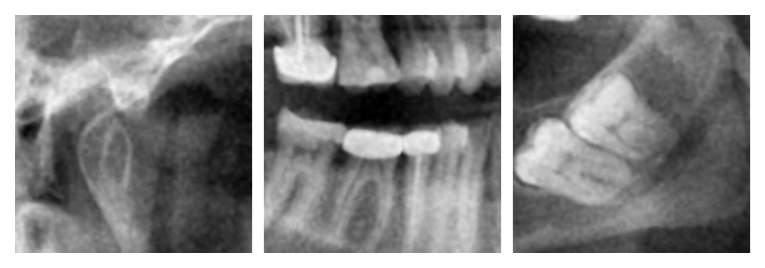}
         \caption{MResDNN}
         \label{fig7d}
     \end{subfigure}
     \begin{subfigure}[b]{0.32\textwidth}
         \centering
         \includegraphics[width=\textwidth]{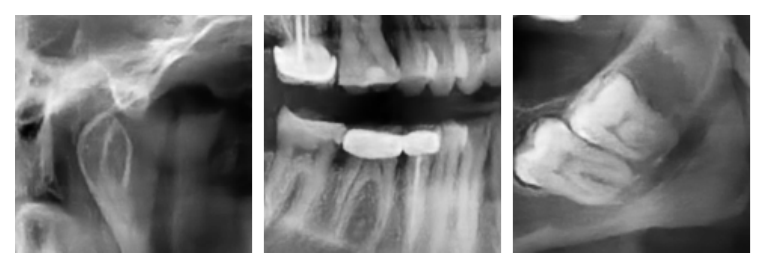}
         \caption{Uformer}
         % \label{fig7e}
     \end{subfigure}
     \begin{subfigure}[b]{0.32\textwidth}
         \centering
         \includegraphics[width=\textwidth]{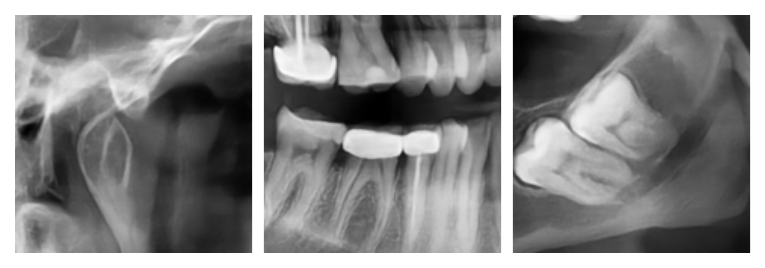}
         \caption{Prop. NAADA}
         % \label{fig7f}
     \end{subfigure}
     \caption{Visual denoising results of the best comparative methods on an unseen noisy image at input PSNR of $14.99$ dB}
    \label{fig7}
\end{figure*}

The results of the testing phase are summarized in Table \ref{tab:testing_results} as average PSNR and SSIM values of all the methods, along with the 95\% confidence interval. Observe from Table \ref{tab:testing_results} that the proposed method outperforms the baseline methods including Uformer, a much more computationally expensive established denoising method. Violin plots for the testing data are shown in Fig. \ref{fig4}. Our approach demonstrates a more concentrated distribution with higher median values for both PSNR and SSIM, indicating that it consistently delivers superior denoising performance with  limited variation. Further observe that the proposed DAE with regular self-attention, termed ADA, is substantially outperformed by the Uformer network, highlighting the impact of the NASA approach in the proposed NAADA method. Moreover, it is interesting to observe that the BM3D yields second highest SSIM values among the SOTA methods (after Uformer), while its PSNR is the lowest. DnCNN turns out to be the most inconsistent method with its SSIM values having highest spread, Fig. \ref{fig4b}.

\subsection{Qualitative evaluation}
For qualitative analysis, we compared the denoised images produced by the proposed method and the baseline methods at three different levels of PSNR, corresponding to low, medium, and high noise.
Furthermore, we also display the zoomed-in views of the specific regions in the denoised images to facilitate visual evaluation. First, we compare the two proposed methods, i.e., NAADA and ADA, in Fig. \ref{fig5} to demonstrate the impact of noise-aware attention in NAADA on the recovered details. Observe that fine anatomical details recovered by NAADA appear much clearer when compared to ADA., indicating the effect of noise-aware attention.

\begin{figure*}
     \centering
     \begin{subfigure}[b]{0.32\textwidth}
         \centering
         \includegraphics[width=\textwidth]{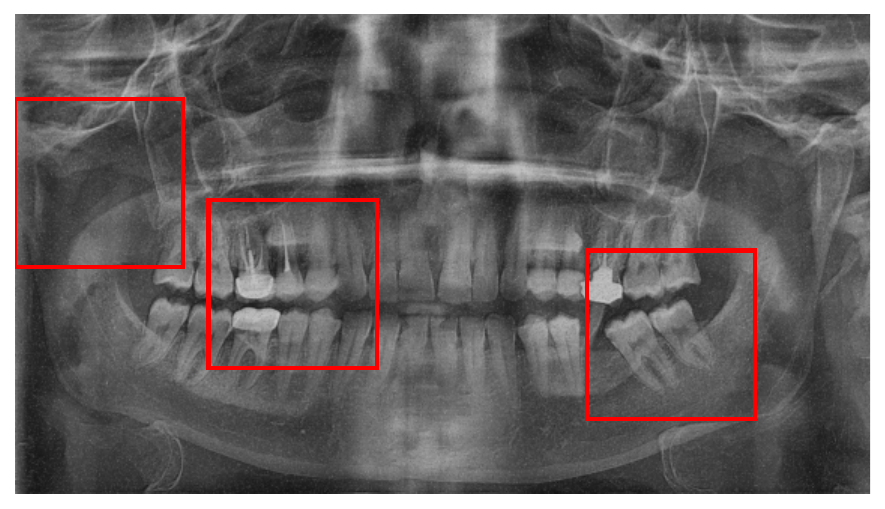}
         % \caption{$y=3\sin x$}
         % \label{fig:three sin x}
     \end{subfigure}
     \begin{subfigure}[b]{0.32\textwidth}
         \centering
         \includegraphics[width=\textwidth]{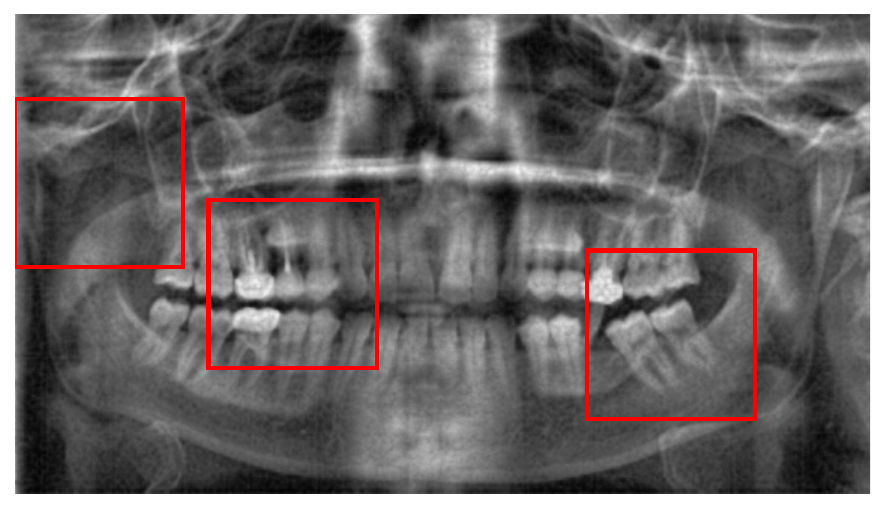}
         % \caption{}
         % \label{fig:y equals x}
     \end{subfigure}
     \begin{subfigure}[b]{0.32\textwidth}
         \centering
         \includegraphics[width=\textwidth]{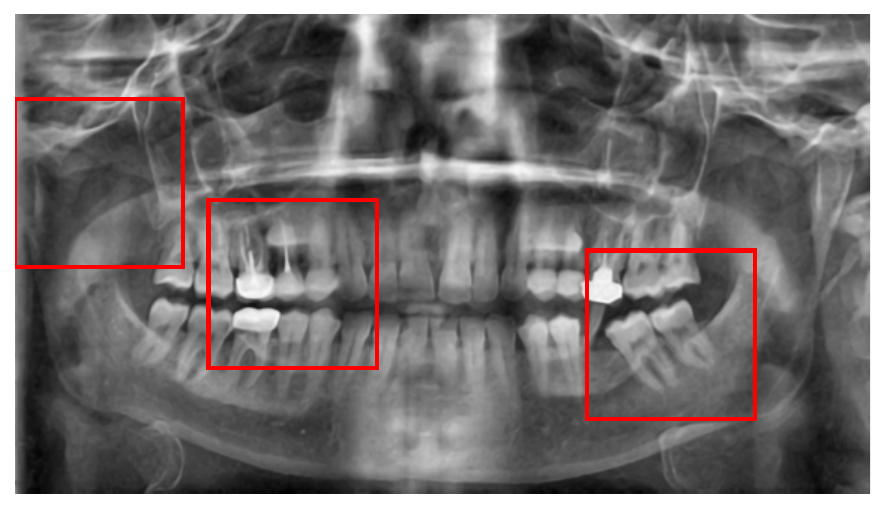}
         % \caption{$y=3\sin x$}
         % \label{fig:three sin x}
     \end{subfigure}

    \begin{subfigure}[b]{0.32\textwidth}
         \centering
         \includegraphics[width=\textwidth]{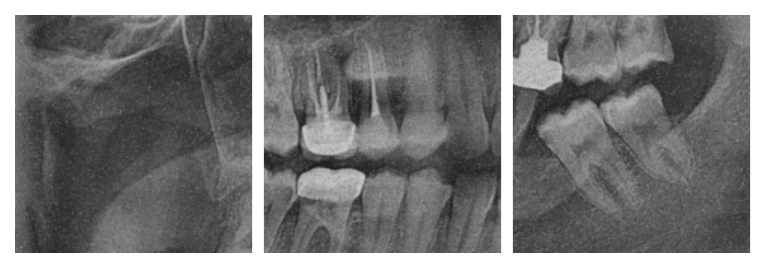}
         \caption{Noisy at PSNR = $18.40$ dB}
         \label{fig8a}
     \end{subfigure}
     \begin{subfigure}[b]{0.32\textwidth}
         \centering
         \includegraphics[width=\textwidth]{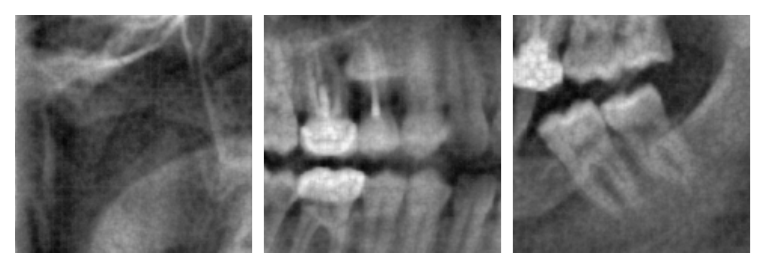}
         \caption{MedDAE}
         \label{fig8b}
     \end{subfigure}
     \begin{subfigure}[b]{0.32\textwidth}
         \centering
         \includegraphics[width=\textwidth]{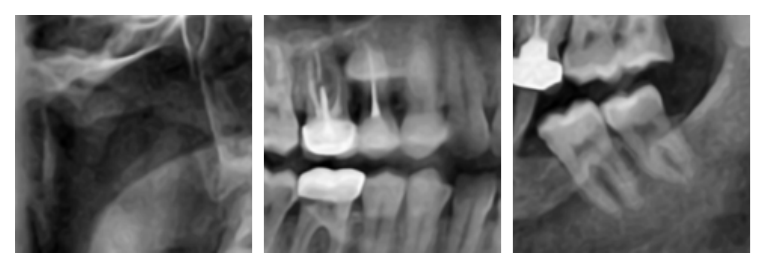}
         \caption{BM3D}
         \label{fig8c}
     \end{subfigure}
     
     \begin{subfigure}[b]{0.32\textwidth}
         \centering
         \includegraphics[width=\textwidth]{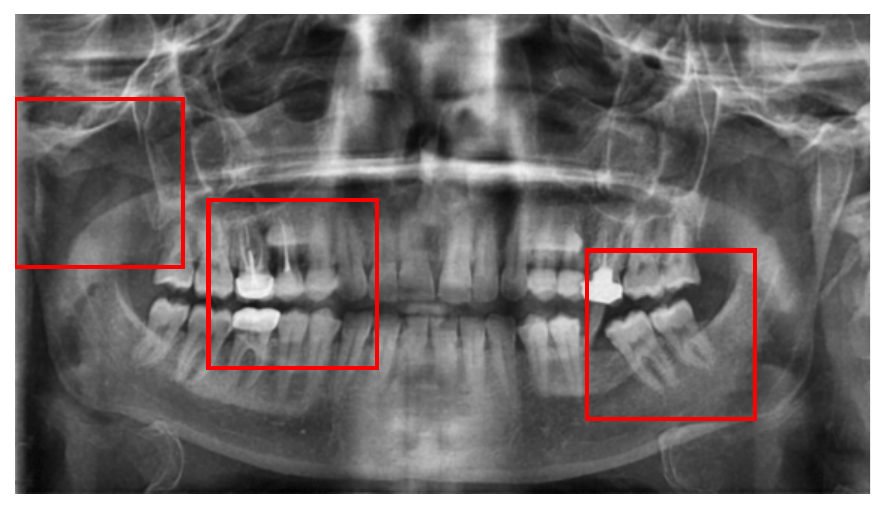}
         % \caption{}
         % \label{fig:y equals x}
     \end{subfigure}
     \begin{subfigure}[b]{0.32\textwidth}
         \centering
         \includegraphics[width=\textwidth]{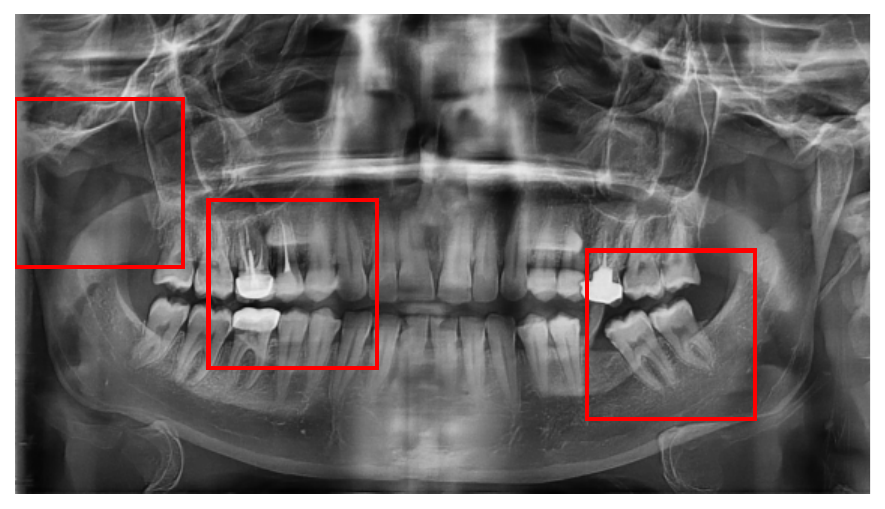}
         %\caption{}
         % \label{fig:three sin x}
     \end{subfigure}
     \begin{subfigure}[b]{0.32\textwidth}
         \centering
         \includegraphics[width=\textwidth]{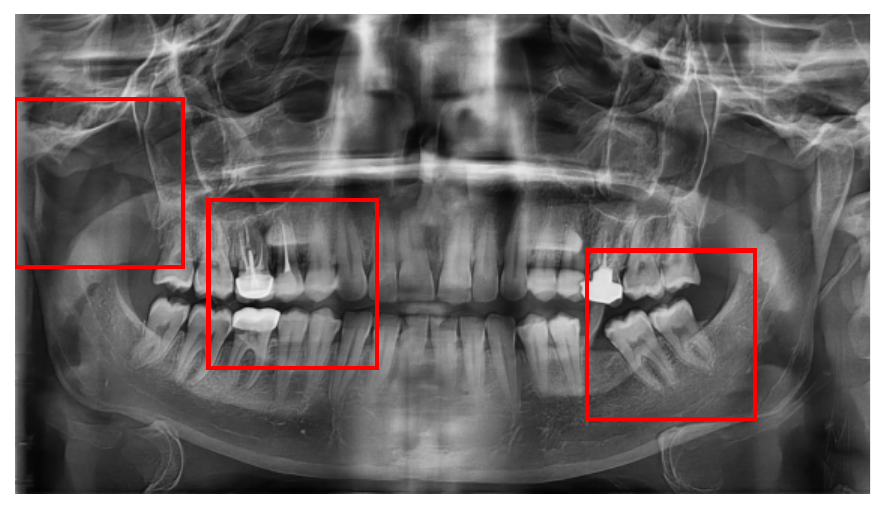}
         % \caption{$y=3\sin x$}
         % \label{fig:three sin x}
     \end{subfigure}
     
     \begin{subfigure}[b]{0.32\textwidth}
         \centering
         \includegraphics[width=\textwidth]{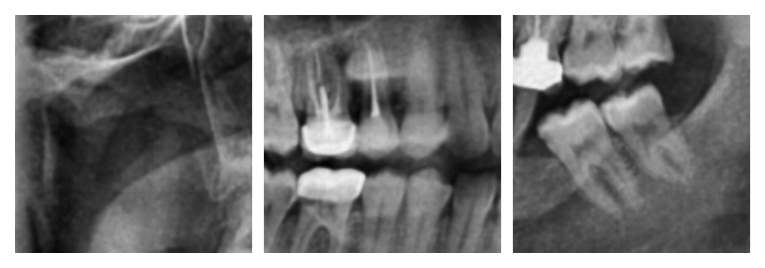}
         \caption{MResDNN}
         \label{fig8d}
     \end{subfigure}
     \begin{subfigure}[b]{0.32\textwidth}
         \centering
         \includegraphics[width=\textwidth]{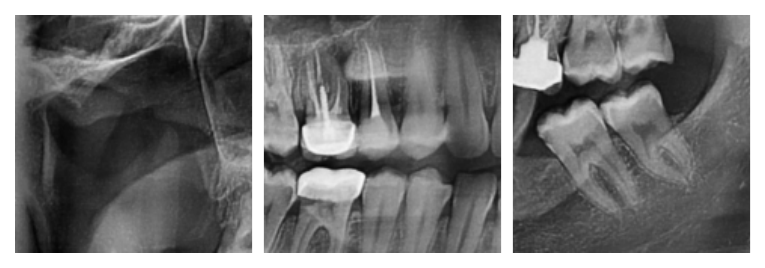}
         \caption{Uformer}
         \label{fig8e}
     \end{subfigure}
     \begin{subfigure}[b]{0.32\textwidth}
         \centering
         \includegraphics[width=\textwidth]{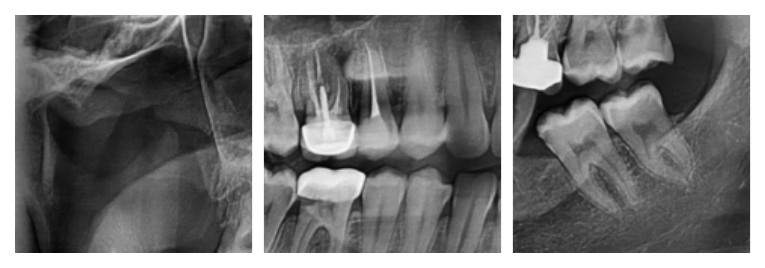}
         \caption{Prop. NAADA}
         \label{fig8f}
     \end{subfigure}
        \caption{Visual denoising results of the best comparative methods on an unseen noisy image at input PSNR of $18.34$ dB}
        \label{fig8}
\end{figure*}

Next, we compare the visual quality of the denoised images produced by the proposed NAADA method against several SOTA approaches, including Uformer, MResDNN, Med. DAE, and BM3D. The DnCNN method was excluded from these visual comparisons due to its inconsistent performance across the testing dataset, as shown by the violin plots in Fig.~\ref{fig4}. We consider three distinct cases of input PSNR: low ($11.74$~dB), medium ($14.99$~dB), and high ($18.34$~dB), as presented in Figs.~\ref{fig6}, \ref{fig7}, and \ref{fig8}, respectively.  
For the case of low PSNR (i.e., higher noise level), shown in Fig.~\ref{fig6}, traditional methods struggled to effectively suppress the noise, often introducing visible blurring or residual artifacts. In contrast, our method managed to reduce noise while preserving finer details. For the medium PSNR case in Fig. \ref{fig7}, the proposed method continues to maintain a balance between noise suppression and structural detail, whereas some comparative methods resulted in over-smoothed images. At high PSNR (low noise levels), shown in \ref{fig8}, all methods performed relatively well, but our approach showed superior performance in maintaining the sharpness and clarity of the image, with minimal noise remnants. These visual comparisons at various noise levels further validate the robustness and effectiveness of our method across a range of image conditions.

\subsection{Clinical evaluation}
We conducted a clinical evaluation of image quality by the aforementioned prosthetic dentist. A total of 50  noisy test images, randomly generated at varying input SNRs using the noise model described in Section~\ref{sec:IIIA}, were denoised using the comparative methods, including the proposed NAADA method. In a blinded randomized forced-choice assessment, the clinical expert was asked to select the image with the best perceptual quality for each case, focusing on anatomical depiction. The results showed that the proposed NAADA and ADA methods were consistently preferred, with NAADA being selected as the best-quality image in 90\% of the cases, and ADA in the remaining 10\%.

\section{Conclusion}
\label{sec:V}
This article introduces NAADA, a Noise-Aware Attention Denoising Autoencoder, that was trained for denoising dental panoramic radiographs. The novel contribution of our work involves a noise-aware attention mechanisms that enhances the proposed denoising autoencoder to effectively emphasize fine anatomical details in the noisy and relatively cleaner regions. Experimental results demonstrate that NAADA outperforms existing denoising approaches in terms of PSNR, SSIM, and perceptual quality, particularly at lower signal-to-noise ratios. Our findings highlight the importance of attention-driven denoising strategies in dental imaging, where maintaining diagnostic integrity is crucial. The proposed model generalizes well across varying noise levels and can be extended to other medical imaging modalities with similar noise characteristics. Future work will explore optimizing NAADA for real-time clinical applications and investigating its integration with downstream tasks such as segmentation and diagnosis.

\section*{Acknowledgment}
\small{This research was funded by the Independent Research Fund Denmark, project “Synthetic Dental Radiography using Generative Artificial Intelligence”, grant ID 10.46540/3165-00237B.}

\bibliographystyle{IEEEtran}
\bibliography{bibtex/bib/IEEEexample}

\begin{thebibliography}{10}
\providecommand{\url}[1]{#1}
\csname url@samestyle\endcsname
\providecommand{\newblock}{\relax}
\providecommand{\bibinfo}[2]{#2}
\providecommand{\BIBentrySTDinterwordspacing}{\spaceskip=0pt\relax}
\providecommand{\BIBentryALTinterwordstretchfactor}{4}
\providecommand{\BIBentryALTinterwordspacing}{\spaceskip=\fontdimen2\font plus
\BIBentryALTinterwordstretchfactor\fontdimen3\font minus
  \fontdimen4\font\relax}
\providecommand{\BIBforeignlanguage}[2]{{%
\expandafter\ifx\csname l@#1\endcsname\relax
\typeout{** WARNING: IEEEtran.bst: No hyphenation pattern has been}%
\typeout{** loaded for the language `#1'. Using the pattern for}%
\typeout{** the default language instead.}%
\else
\language=\csname l@#1\endcsname
\fi
#2}}
\providecommand{\BIBdecl}{\relax}
\BIBdecl

\bibitem{farman2007panoramic_chap1}
A.~G. Farman, ``Panoramic radiology,'' \emph{Getting the most out of Panoramic
  Radiographic Interpretation. Berlin, Heidelberg: Springer-Verlag}, pp. 1--14,
  2007.

\bibitem{fuentes2021panoramic}
R.~Fuentes, A.~Arias, and E.~Borie-Echevarria, ``Panoramic radiographs: an
  invaluable tool for the study of bone and teeth components in the
  maxillofacial region,'' \emph{Int. J. Morphol}, vol.~39, no.~1, pp. 268--73,
  2021.

\bibitem{goebel2005noise}
P.~M. Goebel, A.~N. Belbachir, and M.~Truppe, ``Noise estimation in panoramic
  x-ray images: An application analysis approach,'' in \emph{IEEE/SP 13th
  Workshop on Statistical Signal Processing, 2005}.\hskip 1em plus 0.5em minus
  0.4em\relax IEEE, 2005, pp. 996--1001.

\bibitem{arnold1984noise}
B.~A. Arnold and P.~Scheibe, ``Noise analysis of a digital radiography
  system,'' \emph{American journal of roentgenology}, vol. 142, no.~3, pp.
  609--613, 1984.

\bibitem{abramova2020analysis}
V.~Abramova, S.~Krivenko, V.~Lukin, and O.~Krylova, ``Analysis of noise
  properties in dental images,'' in \emph{2020 IEEE 40th International
  Conference on Electronics and Nanotechnology (ELNANO)}.\hskip 1em plus 0.5em
  minus 0.4em\relax IEEE, 2020, pp. 511--515.

\bibitem{omar2016quantitative}
G.~Omar, Z.~Abdelsalam, and W.~Hamed, ``Quantitative analysis of metallic
  artifacts caused by dental metallic restorations: comparison between four
  cbct scanners,'' \emph{Future Dental Journal}, vol.~2, no.~1, pp. 15--21,
  2016.

\bibitem{dhillon2012positioning}
M.~Dhillon, S.~M. Raju, S.~Verma, D.~Tomar, R.~S. Mohan, M.~Lakhanpal, and
  B.~Krishnamoorthy, ``Positioning errors and quality assessment in panoramic
  radiography,'' \emph{Imaging science in dentistry}, vol.~42, no.~4, pp.
  207--212, 2012.

\bibitem{Abdulbadea2023Enhancing}
O.~E. E.-D. Abdulbadea~Altukroni and S.~Jabeen, ``Enhancing the quality of
  dental radiographic images: A review on panoramic and periapical radiograph
  enhancement techniques,'' in \emph{Advances in Dentistry \& Oral
  Health}.\hskip 1em plus 0.5em minus 0.4em\relax Juniper Publishers, 1999,
  vol.~16, no.~4, pp. 1--12.

\bibitem{goreke2023novel}
V.~G{\"o}reke, ``A novel method based on wiener filter for denoising poisson
  noise from medical x-ray images,'' \emph{Biomedical Signal Processing and
  Control}, vol.~79, p. 104031, 2023.

\bibitem{tacs2023application}
{\.I}.~{\c{C}}. Ta{\c{s}}, ``Application of panoramic dental x-ray images
  denoising,'' \emph{International Journal of Innovative Engineering
  Applications}, vol.~7, no.~1, pp. 13--20, 2023.

\bibitem{dabov2007image}
K.~Dabov, A.~Foi, V.~Katkovnik, and K.~Egiazarian, ``Image denoising by sparse
  3-d transform-domain collaborative filtering,'' \emph{IEEE Transactions on
  image processing}, vol.~16, no.~8, pp. 2080--2095, 2007.

\bibitem{goyal2020bm3d}
B.~Goyal, A.~Dogra, and A.~Sharma, ``Bm3d outperforms major benchmarks in
  denoising: an argument in favor,'' \emph{Journal of Computer Science},
  vol.~16, no.~6, pp. 838--847, 2020.

\bibitem{MLPBM3D2012burger}
H.~C. Burger, C.~J. Schuler, and S.~Harmeling, ``Image denoising: Can plain
  neural networks compete with bm3d?'' in \emph{2012 IEEE conference on
  computer vision and pattern recognition}.\hskip 1em plus 0.5em minus
  0.4em\relax IEEE, 2012, pp. 2392--2399.

\bibitem{DenoisingReview2020tian}
C.~Tian, L.~Fei, W.~Zheng, Y.~Xu, W.~Zuo, and C.-W. Lin, ``Deep learning on
  image denoising: An overview,'' \emph{Neural Networks}, vol. 131, pp.
  251--275, 2020.

\bibitem{ResLearnDen2017zhang}
K.~Zhang, W.~Zuo, Y.~Chen, D.~Meng, and L.~Zhang, ``Beyond a gaussian denoiser:
  Residual learning of deep cnn for image denoising,'' \emph{IEEE transactions
  on image processing}, vol.~26, no.~7, pp. 3142--3155, 2017.

\bibitem{MResDNN2024mittal}
A.~Mittal, N.~Kaur, A.~Gupta, and G.~Singh, ``Deep residual learning-based
  denoiser for medical x-ray images,'' \emph{Evolving Systems}, vol.~15, no.~6,
  pp. 2339--2353, 2024.

\bibitem{hwang2018inception}
S.~Hwang, G.~Yu, H.~T. Nguyen, N.~Shahid, D.~Sin, J.~Kim, and S.~Na,
  ``Inception-residual block based neural network for thermal image
  denoising,'' \emph{arXiv preprint arXiv:1810.13169}, 2018.

\bibitem{shi2019hierarchical}
W.~Shi, F.~Jiang, S.~Zhang, R.~Wang, D.~Zhao, and H.~Zhou, ``Hierarchical
  residual learning for image denoising,'' \emph{Signal Processing: Image
  Communication}, vol.~76, pp. 243--251, 2019.

\bibitem{chen2023DAEreview}
S.~Chen and W.~Guo, ``Auto-encoders in deep learning—a review with new
  perspectives,'' \emph{Mathematics}, vol.~11, no.~8, p. 1777, 2023.

\bibitem{jhamb2018attentive}
Y.~Jhamb, T.~Ebesu, and Y.~Fang, ``Attentive contextual denoising autoencoder
  for recommendation,'' in \emph{Proceedings of the 2018 ACM SIGIR
  international conference on theory of information retrieval}, 2018, pp.
  27--34.

\bibitem{singh2022attention}
P.~Singh and A.~Sharma, ``Attention-based convolutional denoising autoencoder
  for two-lead ecg denoising and arrhythmia classification,'' \emph{IEEE
  Transactions on Instrumentation and Measurement}, vol.~71, pp. 1--10, 2022.

\bibitem{vaswani2017attention}
A.~Vaswani, ``Attention is all you need,'' \emph{Advances in Neural Information
  Processing Systems}, 2017.

\bibitem{zhang2019spatialselfattention}
H.~Zhang, I.~Goodfellow, D.~Metaxas, and A.~Odena, ``Self-attention generative
  adversarial networks,'' in \emph{International conference on machine
  learning}.\hskip 1em plus 0.5em minus 0.4em\relax PMLR, 2019, pp. 7354--7363.

\bibitem{xie2023attentionSegReview}
Y.~Xie, B.~Yang, Q.~Guan, J.~Zhang, Q.~Wu, and Y.~Xia, ``Attention mechanisms
  in medical image segmentation: A survey,'' \emph{arXiv preprint
  arXiv:2305.17937}, 2023.

\bibitem{niu2021reviewAttention}
Z.~Niu, G.~Zhong, and H.~Yu, ``A review on the attention mechanism of deep
  learning,'' \emph{Neurocomputing}, vol. 452, pp. 48--62, 2021.

\bibitem{vincent2008DAEasrobustfeatureextrators}
P.~Vincent, H.~Larochelle, Y.~Bengio, and P.-A. Manzagol, ``Extracting and
  composing robust features with denoising autoencoders,'' in \emph{Proceedings
  of the 25th international conference on Machine learning}, 2008, pp.
  1096--1103.

\bibitem{tihon2021daema}
S.~Tihon, M.~U. Javaid, D.~Fourure, N.~Posocco, and T.~Peel, ``Daema: Denoising
  autoencoder with mask attention,'' in \emph{International Conference on
  Artificial Neural Networks}.\hskip 1em plus 0.5em minus 0.4em\relax Springer,
  2021, pp. 229--240.

\bibitem{lee2018poisson}
S.~Lee, M.~S. Lee, and M.~G. Kang, ``Poisson--gaussian noise analysis and
  estimation for low-dose x-ray images in the nsct domain,'' \emph{Sensors},
  vol.~18, no.~4, p. 1019, 2018.

\bibitem{kalivas1999modeling}
N.~Kalivas, I.~Kandarakis, D.~Cavouras, L.~Costaridou, C.~Nomicos, and
  G.~Panayiotakis, ``Modeling quantum noise of phosphors used in medical x-ray
  imaging detectors,'' \emph{Nuclear Instruments and Methods in Physics
  Research Section A: Accelerators, Spectrometers, Detectors and Associated
  Equipment}, vol. 430, no. 2-3, pp. 559--569, 1999.

\bibitem{kasap2011amorphous}
S.~Kasap, J.~B. Frey, G.~Belev, O.~Tousignant, H.~Mani, J.~Greenspan,
  L.~Laperriere, O.~Bubon, A.~Reznik, G.~DeCrescenzo \emph{et~al.}, ``Amorphous
  and polycrystalline photoconductors for direct conversion flat panel x-ray
  image sensors,'' \emph{Sensors}, vol.~11, pp. 5112--5157, 2011.

\bibitem{hamamci2023dentex}
I.~E. Hamamci, S.~Er, E.~Simsar, A.~E. Yuksel, S.~Gultekin, S.~D. Ozdemir,
  K.~Yang, H.~B. Li, S.~Pati, B.~Stadlinger \emph{et~al.}, ``Dentex: An
  abnormal tooth detection with dental enumeration and diagnosis benchmark for
  panoramic x-rays,'' \emph{arXiv preprint arXiv:2305.19112}, 2023.

\bibitem{kingma2014adam}
D.~P. Kingma, ``Adam: A method for stochastic optimization,'' \emph{arXiv
  preprint arXiv:1412.6980}, 2014.

\bibitem{gondara2016medical}
L.~Gondara, ``Medical image denoising using convolutional denoising
  autoencoders,'' in \emph{2016 IEEE 16th international conference on data
  mining workshops (ICDMW)}.\hskip 1em plus 0.5em minus 0.4em\relax IEEE, 2016,
  pp. 241--246.

\bibitem{wang2022uformer}
Z.~Wang, X.~Cun, J.~Bao, W.~Zhou, J.~Liu, and H.~Li, ``Uformer: A general
  u-shaped transformer for image restoration,'' in \emph{Proceedings of the
  IEEE/CVF conference on computer vision and pattern recognition}, 2022, pp.
  17\,683--17\,693.

\end{thebibliography}

\end{document}